\documentclass[prd,preprintnumbers,preprint,nofootinbib]{revtex4-1}

\usepackage{amsmath,slashed,latexsym,amsfonts,mathrsfs}
\usepackage{epstopdf}

\usepackage{epsfig,subfigure}

\newcommand{\be}{\begin{equation}}
\newcommand{\ee}{\end{equation}}
\newcommand{\bea}{\begin{eqnarray}}
\newcommand{\eea}{\end{eqnarray}}

\def \nn {\nonumber}
\def\vep{\varepsilon}
\def \ep{\epsilon}

\begin{document}

%%%%%%%%%%%%%%%%%%%%%%%%%%%%%%

\preprint{YITP-SB-13-43}

\title{Coordinate-space singularities of massless gauge theories}

\author{Ozan Erdo\u{g}an}

\email{erdogan@insti.physics.sunysb.edu}

\affiliation{C.N.\ Yang Institute for Theoretical Physics 
 and Department of Physics and Astronomy, Stony Brook University, Stony Brook, New York 11794--3840, USA}

\date{\today}

\begin{abstract}
The structure of singularities in perturbative massless gauge
theories is investigated in coordinate space. The pinch singularities
in coordinate-space integrals occur at configurations of vertices
which have a direct interpretation in terms of physical scattering of
particles in real space-time in the same way as for the loop momenta
in the case of momentum-space singularities. In the analysis of vertex functions
in coordinate space, the well-known factorization into hard, soft, and
jet functions is found. By power-counting arguments, it is found that 
coordinate-space integrals of vertex functions have logarithmic
divergences at worst.

\end{abstract}

\maketitle

\tableofcontents
\newpage

\section{Introduction}

The structure of singularities in perturbative gauge theories has long
been a subject of study for theoretical interest and for
phenomenological applications~\cite{smatrix,Dixon:2008gr}\@. There is
a vast literature on the subject, and most modern analyses are carried
out in momentum space to calculate scattering
amplitudes~\cite{Bern:1994zx,Britto:2005fq,Dixon:2011xs,Britto:2010xq}. 
Calculations involving Wilson lines, however, are often simpler in
coordinate space~\cite{Korchemsky:1992xv,Aybat:2006mz} and
coordinate-space integrals were used for Wilson lines 
in the application of dual conformal
invariance~\cite{Drummond:2007aua}\@. 
It is therefore natural to consider using them for amplitudes as well. 
The purpose of this paper is to
provide a new, coordinate-space analysis of singularities in
perturbation theory applicable to
amplitudes for massless gauge theories.

It is well known that the momentum-space singularities of Feynman integrals in a
generic quantum field theory occur at configurations of internal loop
momenta that have a direct interpretation in terms of physical
scattering of on-shell particles in real
space-time~\cite{Landau:1959fi,Coleman:1965xm}\@. In this
study, we will analyze the origin and structure of these singularities
directly in coordinate space.

Massless gauge theories suffer
from infrared (IR) divergences, which characterize the long-distance
contributions to perturbative predictions, in addition to ultraviolet (UV)
divergences, which can be removed by local counterterms. 
An analysis of infrared divergences in gauge theories
from the point of pinch singularities of Feynman integrals over loop
momenta was given by
Ref.~\cite{Sterman:1978bi}\@. Following~\cite{Date:1982un}, which
dealt with scalar theories, we will show that the
coordinate-space singularities of massless gauge theories have the
same interpretation in terms of physical scattering of particles with
conserved momenta. In contrast to momentum space examples, however, we will see
that collinear singularities are of ultraviolet
nature in coordinate space, and require $D<4$ in dimensional
regularization. This analysis can be applied to a variety of 
field theory objects derived from Green functions, including form
factors and vertex functions, Wilson lines, as well as cut diagrams
for cross sections.

In a detailed analysis of vertex functions in coordinate space, we
will find the factorization into hard, soft, and jet functions familiar from
momentum space analysis~\cite{Collins:1989gx,Sterman:1995fz}\@. In coordinate
space, the soft function is finite when the external points are kept
at finite distances from each other. Therefore, ultraviolet
regularization is needed only for the jets and the hard function. Adapting
the power counting technique developed for momentum space
in~\cite{Sterman:1978bi}, the residues of the lightcone poles of
vertex functions in coordinate space  
will be shown to have logarithmic divergences at worst.

This paper is organized as follows. In section~\ref{analysis}, a
brief, general review of pinch singularities will be followed by a derivation
of conditions for singularities in coordinate-space integrals together
with their physical interpretations. We will also comment on the case
with massive lines in Appendix~\ref{app:mass}\@. In section~\ref{form}, we will
analyze the structure of singularities of vertex functions in coordinate
space, solving the conditions for pinch singularities, first
explicitly at lowest loop order and
then extending the solutions to arbitrary order in perturbation theory. In
section~\ref{power}, we will adapt the power counting
technique developed in~\cite{Sterman:1978bi} to the coordinate-space
vertex functions, and show that divergences are at worst
logarithmic relative to their lowest-order results at higher
orders in perturbation theory. In the last section, we will discuss
the approximations that can be made in the integrand to obtain the leading
singularity. We will describe the ``hard-collinear'' and then the
``soft-collinear'' approximations, which will lead to factorization of jets from
the hard and soft functions. Lastly, we will show that the fermionic vertex
function can be approximated by a Wilson line calculation, by imposing
the conditions for a pinch singularity inside the integrands.

\section{Analysis of singularities}\label{analysis}

This section treats the coordinate-space singularities of Feynman
diagrams in gauge theories. The discussion is in many ways similar to the
momentum-space analysis of
Refs.~\cite{smatrix,hwa-homology,sterman-qft}. The results of this
section will be employed to identify the natural subregions of the
corresponding diagrams in order to study their behavior 
in coordinate space.

We start our analysis with an arbitrary Feynman
integral with massless lines in coordinate space. We work in
$D=4-2\vep$ dimensions using dimensional regularization. For gauge
theories, we employ Feynman gauge. The
integrands in scalar and gauge theories are similar, except that in the
latter case, gauge field vertices have derivatives that act on
attached lines. These derivatives change the powers of denominators and
produce numerator factors, which may enhance or suppress the
integrals.

In coordinate space, we can represent graphical integrals schematically as
\be I(\{x^{\mu}_{i}\})=\prod_{\mathrm{vertices}\ k}\int d^Dy_k
\prod_{\mathrm{lines}\ j}\,
\frac{1}{\left[-(\sum_{k'}\eta_{jk'}\,X_{k'})^2+i\ep\right]^{p_j}}\times F(x_i,y_k,D) \ , \label{master1}
\ee
where the positions of internal vertices $y^{\mu}_k$ are integrated
over all space-time for fixed external points $x^{\mu}_i$\@. For each
line, the sum over $\{X_{k'}\}=\{y_k,x_i\}$ includes all
vertices, internal and external, where $\eta_{jk}$ is an ``incidence matrix'',
which takes values $+1$ and $-1$ when the line $j$ ends or begins at vertex
$k$, respectively, and is zero otherwise.  The orientation
of a line is at this point arbitrary, but we will see that at
singularities it is determined by the time-ordering of the vertices it connects.
Before the action of derivatives, the power of the denominator of line
$j$ is $p_j=2-\vep$ for fermion lines, and $p_j=1-\vep$ for scalar and
gauge field lines; however, if a derivative acts on a line, the power
of its denominator is increased by $1$\@. This
expression holds for scalar and gauge theories, for which we sum over
terms with different numbers of derivatives, and the
functions $F(x_i,y_k,D)$ include remaining constants, group theory
factors, and numerator factors, which do not
affect the locations of the singularities but will matter in power
counting. They are simply numerical constants for scalar
theories. For theories with spin, they also carry the spin-dependence,
which we have suppressed here. The integrand in (\ref{master1}) becomes
singular when a line moves to the lightcone.

After combining the propagators of each
line with Feynman parametrization, the integral will be of the form, 
\be\begin{split} I(\{x^{\mu}_{i}\})= \prod_{\mathrm{lines}\
  j}\int^1_0d\alpha_j&\,\alpha^{p_j-1}_j\,\delta(1-\sum_j\alpha_j) \\
&\times\prod_{\mathrm{vertices}\
k}\int d^Dy_k\,D(\alpha_j,x_i,y_k)^{-N(\vep)}\,\bar{F}(x_i,y_k,D) \
, \label{fgmaster} 
\end{split}\ee
where we have absorbed the prefactors of the parametrization into
$\bar{F}(x_i,y_k,D)$, and where the common denominator is given by
\be 
D(\alpha_j,x_i,y_k)=\sum_j \alpha_j\left[-z^2_j(x_i,y_k)+i\ep\right] \ . \label{denom}\ee
Here, $\alpha_j$ is the Feynman parameter of the
$j$th line, and $z^{\mu}_j$ denotes the argument of its
propagator, which is the coordinate difference between the vertices it
connects. The overall power of the denominator is $N(\vep)=\sum_j
p_j(\vep)$, in particular, $N(\vep)=N(1-\vep)$ for a diagram with $N$
scalar lines only. For gauge theories, for a diagram with $N_g$ gauge field
lines, $N_f$ massless fermion lines, and $V_{3g}$ three-vector vertices
it is given by $N(\vep)=N_f(2-\vep)+N_g(1-\vep)+V_{3g}$\@.

The zeros of the denominator $D(\alpha_j,x_i,y_k)$ in
Eq. (\ref{denom}) determine the
positions of the poles of the integrand in (\ref{fgmaster}). These
poles may produce branch points of
$I(\{x^{\mu}_{i}\})$, depending on whether or not they may be avoided
by contour deformation in the complex
$(\alpha, y)$-space. We recall here the summary given
in~Ref.~\cite{smatrix}\@. In general, the singularities of a function
$f(z)$ defined by a single integral,
\be f(z)\ =\ \int_{\mathcal{C}} dw\,\frac{1}{g(z,w)} \ , \ee
arise if and only if
the poles $\tilde{w}(z)$ of the integrand, which are zeros of $g(z,w)$,
cannot be avoided by contour deformation. This follows from a theorem proven by
Hadamard~\cite{Hadamard}, and happens either when one
of the poles migrates to one of
the end-points of the contour, an {\em end-point singularity} or when two or
more isolated poles coalesce at a point trapping the contour
between them, resulting in a {\em pinch singularity}.

These conditions for the existence of singularities can be generalized
as necessary conditions for functions of several (external) variables
that are defined by multiple integrals,
\be f(\{z_i\})\ =\ \int_{\mathcal{H}}
\prod_jdw_j\,\frac{1}{g(\{z_i\},\{w_j\})} \ , \ee
such as $I(\{x^{\mu}_{i}\})$ in our case. Here, the
hypercontour~$\mathcal{H}$ denotes the multi-dimensional region of
integration. The set of points $S=\{\tilde{w}(\tilde{z})\}$ on which
$g(\{z_i\},\{w_j\})=0$ defines surfaces in the complex
$(z,w)$-space. If $g(\{z_i\},\{w_j\})$ factors as $g=g_1(\{z_i\},\{w_j\})\times\cdots\times
g_r(\{z_i\},\{w_j\})$, then there are $r$~such singular surfaces,
which may or may not intersect with each other. 
As in the case of an integral over a single variable, the
singularities occur when an intersection of these singular surfaces with
the hypercontour~$\mathcal{H}$ cannot be avoided. Summarizing the
arguments presented in Ref.~\cite{smatrix}, this again happens
either when
a singular surface $S$ overlaps with the boundary of $\mathcal{H}$ ({\em
  end-point singularity}) or when the
hypercontour~$\mathcal{H}$ is trapped between two or more singular
surfaces or between two different parts of the same singular surface 
({\em pinch singularity}). At an end-point, $\mathcal{H}$~cannot be
moved in the directions normal to its boundary, while at a pinch it cannot be
moved away from singular regions in the direction of the normals to
two (or more) singular surfaces, which are in opposite directions. 
In both cases, the vanishing of the gradient of $g(\{z_i\},\{w_j\})$
on $S$ is the
necessary condition,
\be \left.\frac{\partial}{\partial w_j}\,g(\{z_i\},\{w_j\})\right|_{g=0}\ =\ 0 \ .\ee
In the following, we use the terminology of Ref.~\cite{sterman-qft},
and call the variables that parametrize directions out of the singular surface
$S$ {\em normal}, and those that lie in the surface {\em
  intrinsic}\@. The larger the volume of normal space, the
less singular the integral. 
Refs.~\cite{smatrix,hwa-homology,sterman-qft} present pedagogical
discussions of these concepts.

This reasoning enables us to derive a powerful set of necessary
conditions for singularities of integrals like $I(\{x^{\mu}_i\})$ in
Eq.~(\ref{master1}) using the representation in Eq.~(\ref{fgmaster}), 
where a singular surface $S$ in~$(\alpha, y)$-space is defined by the
set of points $S=\{\tilde{\alpha},\tilde{y}\}$ on which $D(\alpha_j,x_i,y_k)$
vanishes. 
The singularities of (\ref{fgmaster}) can come only from
the end-point $\alpha_j=0$ of the $\alpha_j$-integral, because
$D(\alpha_j,x_i,y_k)$ is linear in the $\alpha_j$\@. Note that
$\alpha_j=1$ is not a different end-point
singularity, as it sets all $\alpha_i$, $i\neq j$, to zero because of
the delta function. On the other hand, there are no end-point
singularities in $y$-integrals, since they are unbounded. 
However, in $y$-integrals the contour of integration can be trapped at a pinch
singular point when the two solutions of the quadratic equation $D=0$
are equal, {\it i.e.},
\be \left.\frac{\partial}{\partial
    y^{\mu}_k}\,D(\alpha_j,x_i,y_k)\right|_{D(\tilde{\alpha},\tilde{y})=0}\
=\ 0 \ . \label{2dlandau} \ee
The momentum space analogs of these conditions are summarized as the
Landau equations~\cite{Landau:1959fi} in the literature. They were also written in coordinate space for scalar
theories in \cite{Date:1982un}. In coordinate space, they are given by
Eq.~(\ref{2dlandau}) above,
\be \alpha_j\ =\ 0 \ , \quad\mathrm{or}\quad z^2_j\ =\ 0 \ ,\label{landau1}\ee
and
\be \sum_{\mathrm{lines}\ j\ \mathrm{at\ vertex}\
  k}\eta_{kj}\,\alpha_j\,z^{\mu}_j\ =\ 0  
\label{landau2}\ . \ee
The conditions in the first line come from $D=0$, while those in the
second line come from $(\partial /\partial y^{\mu}_k)D=0$\@. The
``or'' in the first line is not necessarily exclusive. The
condition $(\partial /\partial\alpha_j)D=0$ for all $j$ is equivalent to $D=0$
since $D$ is homogenous of degree one in the $\alpha_j$\@.

A physical interpretation of the momentum space Landau equations was
originally given by Coleman and Norton in \cite{Coleman:1965xm}. The
momentum space analog of Eq.~(\ref{landau2}) in terms of momenta
$k^{\mu}_i$ of lines is, 
\be \sum_{\mathrm{lines}\ i\ \mathrm{in\ loop}\
  l}\eta_{li}\,\alpha_i\,k^{\mu}_i\ =\ 0 \ .\ee
Then, 
with the identification of
$\alpha_ik^{\mu}_i\equiv\Delta x^{\mu}_i$ with a space-time vector for
each on-shell line, these relations can be thought as describing
on-shell particles propagating between the end- and starting points of
line $i$, which are separated by interval $\Delta x^{\mu}_i$. This
way, $\alpha_i$ is interpreted as the
ratio of the time of propagation to the energy of particle $i$; and thus the
analog of Eq. (\ref{landau1}) states that there is no propagation for
an off-shell line.

Similarly, in coordinate space, after the rescaling $\Delta\bar{x}^{\mu}_j=\alpha_jz^{\mu}_j$, Eq. (\ref{landau2}) directly gives the
same physical picture of on-shell particles propagating in
space-time. The interpretation with particles
propagating forward in time fixes the orientation of lines by
the time-ordering of vertices. 
Additionally, we may identify the product
$\alpha_jz^{\mu}_j$ with a momentum vector,
\be p^{\mu}_j\ \equiv\ \alpha_j\,z^{\mu}_j \ . \ee
Then Eq. (\ref{landau2}) gives momentum conservation for the on-shell
lines with momenta~$p^{\mu}_j$ flowing in or out of vertex $k$. Moreover, with a
further identification of $\alpha_j$ as the ratio of the energy of
line $j$ to the time component of $z_j$,  
\be \alpha_j \ \equiv\ p^0_j/z^0_j  \ , \label{alphaE}\ee
we obtain a relation between the energies and momenta of the
propagating particles associated with the pinch singularities of
Eq. (\ref{fgmaster}),
\be p^{\mu}_j\ = \ E_j\,v^{\mu}_j \ , \quad\mathrm{with}\
v^{\mu}_j=(1,\vec{z}/z^0_j) \ . \ee
This is the relation between energy and momentum of free massless
particles; the magnitude of their velocity is indeed $c=1$ since $(z^0)^2-|\vec{z}|^2=0$\@. Therefore, to each pinch singularity we can associate a
physical picture in which massless particles propagate freely on the lightcone
between vertices, while their momenta satisfy momentum conservation at
each internal vertex as well~\cite{Date:1982un}\@.

In the physical picture above, only lines on the lightcone ``carry''
finite momenta. Lines not on the
lightcone, that is lines connecting vertices at finite distances, have
$\alpha_j=0$ which by Eq.~(\ref{alphaE}) sets their $p_j^{\mu}=0$. In
momentum space, because the momenta of lines with $\alpha_j=0$ do not
show up in the momentum-space analog of~(\ref{landau2}), in a
graphical representation, one can contract such off-shell lines to
points. The resulting diagrams are called {\em reduced diagrams} that
represent {\em lower-order} singularities of a Feynman diagram, while
the diagram with all the lines on the mass-shell ({\em i.e.} no lines
with $\alpha=0$) is said to give the {\em leading}
singularity~\cite{smatrix,Cachazo:2008vp}\@. In contrast, in coordinate space,
these ``contracted'' lines should be compared to ``zero lines'', with
$z_j^{\mu}=0$ that do not contribute to the sum in (\ref{landau2})
either. They represent ``short-distance'' (UV) singularities, which
occur when two connected vertices coincide at the same point, but
are not {\em lower-order} singularities of the coordinate
integral. These pinch singularities originate from the denominator of
a single propagator, where the contour of integration, the real
line, is pinched between two poles 
of the same propagator. Therefore, we
will first identify such UV singularities of an arbitrary integral
like (\ref{master1}), and then
combine the rest of the denominators by Feynman parametrization to
find other types of singularities from groups of lines in the
remaining integrals using the Landau
conditions~(\ref{landau1}) and (\ref{landau2}). We should note that
not all UV singularities give UV {\it divergences}\@. Divergences can
be identified by the power counting procedure below.

As an example of the
application of Eq.~(\ref{2dlandau}) to coordinate-space
integrals, we shall now find the configurations of lines
for pinches in the integration over the position of a single
three-point vertex at a point $y^{\mu}$\@. 
For simplicity, let us consider the following integral in a scalar
theory
\be I(x_1,x_2,x_3)\ =\ \int d^Dy\, \prod^3_{i=1}\,
\frac{1}{[-(x_i-y)^2+i\ep]^{1-\vep}} \ . \ee  
Apart from the the UV singularities when $y^{\mu}=x^{\mu}_i$ for
$i=1,2\mathrm{\ or\ }3$, the
conditions for a pinch between different lines in the $y^{\mu}$ integral
are given by Eqs.~(\ref{landau1}) and (\ref{landau2}) after Feynman
parametrization, 
\be \alpha_1\,z^{\mu}_1+\alpha_2\,z^{\mu}_2+\alpha_3\,z^{\mu}_3\ =\ 0
\ , \label{3pt}\ee
with $z^{\mu}_i\equiv x^{\mu}_i-y^{\mu}$\@. For a pinch singularity,
these vectors are either lightlike, $z^2_i=0$, or have 
$\alpha_i=0$\@. Equation~(\ref{3pt}) cannot be satisfied if all three
vectors have positive entries. Thus, 
at least one external point must have $x^+_i<y^+$ and one must have
$x_j^+ > y^+$, so that there is at
least one incoming and one outgoing line. These considerations
naturally provide a time ordering for vertices
and a direction for lines at any singularity. Assuming all
$\alpha_i\neq 0$ and that all lines are on the
lightcone, $z^2_i=0$ and Eq.~(\ref{3pt}) imply that $z_j\cdot
z_k=0$ as well. That is, all of these lines are parallel. If any one
of the lines is off the lightcone with $\alpha_i=0$, then the other
two are on the lightcone and again parallel to each other
by (\ref{3pt}). These pinch singularities can be interpreted as
a merging or splitting of three particles, which occurs at
point~$y^{\mu}$, with the ratios of their momenta given by the ratios of
the $\alpha_i$\@. Note that these results hold for the three-point
vertices of a gauge theory as well, and can be generalized to
$n$-point vertices. The coordinate-space singularities
of Green functions represent physical particle scattering and thus
can be related naturally to physical scattering amplitudes.

\section{Coordinate-space singularities at a vertex}\label{form}

We will now study how coordinate-space
singularities in a vertex function in a massless gauge theory emerge from
pinches in Feynman integrals in perturbation theory. For simplicity,
the first example that we consider will be the correlation of two
scalar fields with a color-singlet gauge current. We also discuss the
correlation of fermions with the same kind of current. The results of
Sec.~\ref{analysis} will be applied to identify the configurations
that can lead to singularities of such vertex functions in coordinate space.

The scalar vertex function of interest is obtained from the vacuum
expectation value of the time-ordered product of two charged scalar fields
with an incoming color-singlet current,
\be
\Gamma^{\nu}_{S}(x_1,x_2) \ = \ \left\langle
  0\left|T\left(\Phi(x_2)\,J^{\nu}(0)\,\Phi^*(x_1)\right)
  \right|0\right\rangle \ . \label{svertexfnc}\ee
Here, we have shifted the position of the current to the origin using the
translation invariance of the vacuum state. $\Gamma^{\nu}_{S}(x_1,x_2)$
transforms as a vector under Lorentz transformations. Its functional
form is well-known and is determined by the abelian Ward identity, 
\be -i\left(\partial_1+\partial_2\right)_{\nu}\,\Gamma^{\nu}_S(x_1,x_2)\ =\ 
 \left[\delta^D(x_2)\,-\,\delta^D(x_1)\right]\,G_{2}((x_2-x_1)^2) \
 , \label{abelianWI} \ee
where $G_{2}$ is the scalar two-point function, which is only a
function of the invariant distance between the external points.  A
general solution to this
inhomogenous partial
differential equation can be given by a particular solution that
satisfies~(\ref{abelianWI}) plus the general solution to the
homogenous equation,
\be (\partial_{1}+\partial_{2})_{\mu}\Gamma^{\mu}_{S,(H)}(x_1,x_2)=0 \ . \ee 
A particular solution to the abelian Ward identity, which has the
structure of the lowest order result, is given by
\be 
\Gamma^{\nu}_{S,(I)}(x_1,x_2,\mu)\ =\ \left(
\frac{x^{\nu}_2}{(-x^2_2+i\ep)^{1-\vep}}-\frac{x^{\nu}_1}{(-x^2_1+i\ep)^{1-\vep}}\right)\frac{\Sigma_S\big(\mu^2(x_2-x_1)^2\big)}{x^2_1\,x^2_2}
\ , \label{inhscalar}  \ee  
where the form factor $\Sigma_S(\mu^2(x_2-x_1)^2)$ is a
dimensionless function, with
$\mu^2$ the renormalization scale, and is
related to the renormalized scalar two-point function by
\be i\,G_{2}(x^2,\mu^2)\ =\
\frac{\Sigma_S(\mu^2x^2)}{(-x^2+i\ep)^{1-\vep}} \ . \ee
Note that at zeroth order one obtains $\Sigma^{(0)}(x_1,x_2)=1$ from
both equations above.

The general solution to the homogenous equation
can be found easily in momentum space, since one then
has an algebraic equation,
\be (p^{\mu}_1-p^{\mu}_2)\tilde{\Gamma}^{\mu}_{S,(H)}(p_1,p_2)=0 \ , \ee 
whose solution
involves polynomials of momenta times one independent
function\footnote{Starting from $(p_1-p_2)_{\mu}( A\,p^{\mu}_1 +
  B\,p^{\mu}_2)\ = 0 $, the homogenous equation is solved for 
$\frac{A}{B} = \frac{p^2_2-p_1\cdot p_2}{p^2_1-p_1\cdot p_2}$\@. The
Ward identity reduces the number of independent functions by
one.}\@. Here, momentum $p_1$ flows into the vertex and $p_2$ out. The
general solution in momentum representation is given by
\be \tilde{\Gamma}^{\mu}_{S,(H)}(p_1,p_2)=\Big[(p^2_2-p_1\cdot
  p_2)\,p^{\mu}_1 + (p^2_1-p_1\cdot
  p_2)\,p^{\mu}_2\Big]\,\tilde{f}_{H}(p_1,p_2)\ . \ee
After inverse Fourier transform with $p^{\mu}_1\rightarrow
i\partial^{\mu}_1$ and $p^{\mu}_2\rightarrow -i\partial^{\mu}_2$, the
part of the vertex that vanishes in the abelian Ward
identity~(\ref{abelianWI}) is of the form,
\be \Gamma^{\mu}_{S,(H)}(x_1,x_2)\ = \
-i(\partial_1-\partial_2)_{\nu}[\partial^{\mu}_1\partial^{\nu}_2
- \partial^{\nu}_1\partial^{\mu}_2]\,f_{H}(x_1,x_2) \ ,\label{homscalar} \ee
where $f_H$ is a function of mass dimension two. In conventional terms, the
inhomogenous solution gives the `longitudinal' part of the vertex
while the homogenous solution is the `transverse' part. Note that any $f_H$
that is a
function of only $(x_1\pm x_2)^2$ vanishes under the
derivatives in Eq.~(\ref{homscalar})\@. Thus, $f_H$ must depend on
$x^2_i$ separately to contribute to the scalar vertex. This will allow lightcone
singularities to factorize from the rest of the vertex.

The fermionic counterpart of Eq.~(\ref{svertexfnc}) is,
\be 
\left(\Gamma_{F}\right)^{\nu}_{ba}(x_1,x_2) \ = \ \left\langle
  0\left|T\left(\psi_{b}(x_2)\,J^{\nu}(0)\,\bar{\psi}_{a}(x_1)\right)
  \right|0\right\rangle \ , \label{fvertexfnc}
\ee
whose tensor and Dirac structure is determined by the invariance under the
global symmetries of the theory while its functional form is similarly
constrained by the Ward identity for fermion fields. Chiral
invariance for a massless theory requires this vertex to have odd
number of gamma matrices. Skipping the details given for scalars above,  
a particular solution for the `longitudinal' part of the fermionic
vertex function is given by 
\be 
\left(\Gamma_{F,(I)}\right)^{\nu}_{ba}(x_1,x_2,\mu)\ =\
\frac{(\slashed{x}_{2}\,\gamma^{\nu}\,\slashed{x}_{1})_{ba}}{(-x^2_2+i\ep)^{2-\vep}\,(-x^2_1+i\ep)^{2-\vep}}\,\Sigma_{F}\big(\mu^2(x_2-x_1)^2\big) \ ,  \label{inhfermion}
\ee  
where $\Sigma_F(x^2)$ is related to the renormalized fermion
two-point function~$S_F(x^2)$ by
\be  i\,S_{F}(x^2,\mu^2)\ =\
\slashed{x}\,\frac{\Sigma_F(\mu^2x^2)}{(-x^2+i\ep)^{2-\vep}} \ . \ee 
The `transverse' part of the
fermionic vertex function that vanishes in the abelian Ward
identity can be written in the form,
\bea
\Gamma^{\mu}_{F,(H)}(x_1,x_2) & = &
\gamma^{\mu}\left[(\Box_1+2\partial_1\cdot\partial_2+\Box_2)f_1(x_1,x_2)
+(\Box_1-\Box_2)f_2(x_1,x_2)\right] \nonumber\\
 & & -(\slashed{\partial}_1+\slashed{\partial}_2)\,(\partial_1+\partial_2)^{\mu}\,f_1(x_1,x_2)
 \nonumber \\
& &
-
(\slashed{\partial}_1+\slashed{\partial}_2)\,(\partial_1-\partial_2)^{\mu}\,f_2(x_1,x_2) \label{homfermion}
\\
& & +(\slashed{\partial}_1-\slashed{\partial}_2)\,(\partial_1+\partial_2)_{\nu}\,\big(\partial^{\nu}_2\,\partial^{\mu}_1
- \partial^{\nu}_1\,\partial^{\mu}_2\big)\,f_3(x_1,x_2)  \nonumber \\
 & &
 +\gamma_5\,\ep^{\mu\nu\rho\sigma}\,\gamma_{\nu}\,(\partial_1-\partial_2)_{\rho}\,(\partial_1+\partial_2)_{\sigma}\,f_4(x_1,x_2)
 \nonumber \ . 
\eea
Here, all form factors $f_i(x_1,x_2)$ have mass dimension four except
$f_3(x_1,x_2)$, which has dimension two. The tensor
decomposition of this vertex and the list of form factors in momentum
space can be found in Ref.~\cite{Ball:1980ay}\@. Again, these form factors can
have arbitrary dependence on $x^2_i$, which allows factorization of
lightcone singularities.

We are interested
in singularities that are related to scattering processes, thus the
limit $x_1\rightarrow x_2$ will not be considered in the
discussion below as it gives effectively a two-point function. We also
assume here
that $x^{\mu}_1$ and $x^{\mu}_2$ are fixed, nonzero vectors that are 
not lightlike separated~($x_1\cdot x_2\neq 0$)\@. 
Given these external data, the only power singularities of the
coordinate-space vertex functions will be in~$x^2_i$, which correspond
to single-particle poles of the external propagators in momentum
space. Furthermore,
both $\Sigma_S(x_1,x_2)$ in (\ref{inhscalar})
and $\Sigma_F(x_1,x_2)$ in~(\ref{inhfermion}) remain
finite when both $x^2_i=0$ with $x_1\cdot x_2\neq 0$\@. Thus, the leading
divergence of the scalar vertex can come from $f_H(x_1,x_2)$
in~(\ref{homscalar}) while the leading divergence of the fermionic
vertex comes along again from the `transverse' part of the vertex.

\begin{figure}[t]
\centering
\includegraphics[scale=1]{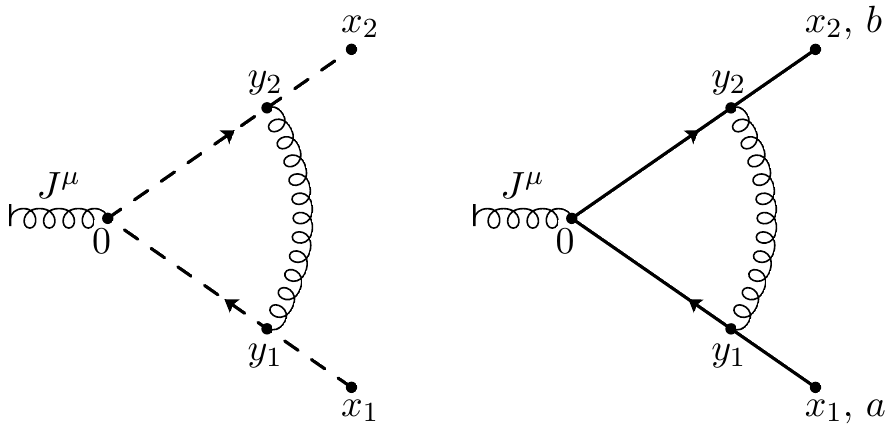}
\caption{One-loop diagrams for the vertex functions in
  Eq.~(\ref{svertexfnc}) and Eq.~(\ref{fvertexfnc}).}
\label{fig:1loopvertex}
\end{figure}

Let us now illustrate how the lightcone singularities of the vertex
functions given above emerge at one loop in perturbation theory. The integrand
of the one-loop diagram in
Fig.~\ref{fig:1loopvertex}, the first nontrivial contribution to
the scalar vertex function, is of the form, 
\be\begin{split} 
\Gamma^{(1)\,\mu}_{S}(x_1,x_2)\ & = \ 
C^{(1)}_S\int 
d^Dy_1\,d^Dy_2\,g^{\alpha\beta}\,\frac{1}{[-(y_2-y_1)^2+i\ep]^{1-\vep}}\,
\\
& \times \left(
\frac{1}{[-(x_2-y_2)^2+i\ep]^{1-\vep}}\,\overleftrightarrow{\partial_{y^{\alpha}_2}}\,
 \left[\frac{1}{[-(y_2-z)^2+i\ep]^{1-\vep}}\,\overleftrightarrow{\partial_{z^{\mu}}}\right.\right. \\
&\quad\quad
\left.\frac{1}{[-(z-y_1)^2+i\ep]^{1-\vep}}\right]_{z=0}
\left.\,\overleftrightarrow{\partial_{y^{\beta}_1}}\,\frac{1}{[-(y_1-x_1)^2+i\ep]^{1-\vep}}\right)
\ ,  
\end{split}\label{1loopvrtxS}\ee
with $C^{(1)}_S$ a numerical constant. Compare this
expression to that of the one-loop diagram for the fermionic vertex function,
\be\begin{split}
\Gamma^{(1)\,\mu}_{F}(x_1,x_2)\ =\  & 
C^{(1)}_F\int 
d^Dy_1\,d^Dy_2\,\frac{1}{[-(y_2-y_1)^2+i\ep]^{1-\vep}}\ \\
 &
 \times\left(\slashed{\partial}_{x_2}\frac{1}{[-(x_2-y_2)^2+i\ep]^{1-\vep}}\right)\gamma^{\alpha}\left(\slashed{\partial}_{y_2}\frac{1}{[-y_2^2+i\ep]^{1-\vep}}\right)\gamma^{\mu}
 \\
&\times \left(\slashed{\partial}_{y_1}\frac{1}{[-y_1^2+i\ep]^{1-\vep}}\right)\gamma_{\alpha}\left(\slashed{\partial}_{x_1}\frac{1}{[-(y_1-x_1)^2+i\ep]^{1-\vep}}\right)\ . 
\end{split} \label{1loopvrtxF}\ee
Clearly, both have the same pole structure, more precisely, the
positions of the poles are the same, although term-by-term their
degrees may or may not be different. Using Feynman parametrization,
either before or after the action of the derivatives on the lines,
both integrals can be put into the form of Eq.~(\ref{fgmaster}) with
the same common denominator but, of course, with different numerator
factors and different powers of the resulting denominator in the integrands.

The pinch singularities of (\ref{1loopvrtxS}) and (\ref{1loopvrtxF}) can
originate from the two poles of a single denominator or from poles from
different denominators. 
For the latter case, we will combine the denominators by Feynman
parametrization, and use Eq.~(\ref{2dlandau}), the
coordinate-space analog of Landau equations. In general, at a pinch
singularity, one
encounters combinations of these two cases simultaneously. After
Feynman parametrization the common
denominator for either diagram is given by
\be\begin{split}
D^{(1)}\big(x_1,x_2,y_1,y_2;\{\alpha\}\big)\ =\
&-\alpha_1\,(y_1-x_1)^2-\alpha_2\,y^2_1-\alpha_3\,y^2_2 \\
 & -\alpha_4\,(x_2-y_2)^2-\alpha_5\,(y_2-y_1)^2+i\ep \
 . \end{split}\ee
Using Eq.~(\ref{2dlandau}), we get the Landau conditions
for pinches in the integration over the positions of internal vertices $y_1$ and
$y_2$,
\bea
\alpha_1\,(y_1-x_1)^{\mu}+\alpha_2\,y^{\mu}_1-\alpha_5\,(y_2-y_1)^{\mu}&=&
0 \ , \label{lan11} \\
-\alpha_4\,(x_2-y_2)^{\mu}+\alpha_3\,y^{\mu}_2+\alpha_5\,(y_2-y_1)^{\mu}&=&
0 \ , \label{lan12}
\eea
when $D^{(1)}$ vanishes at a singularity.

We shall list the singularities of (\ref{1loopvrtxS}) or
(\ref{1loopvrtxF}) starting with pinches without any zero
lines. Eqs.~(\ref{lan11}) and (\ref{lan12}) make up an underdetermined
system of two vector equations, since the $\alpha_j$
also are unknowns. 
We will not list all solutions to these equations
but only those that give the leading power singularities of the vertex
functions, which are physically relevant. 
By leading power singularities of the vertex functions we mean the
terms in
\be \hat{\Gamma}^{\mu}_{S,F}(x_1,x_2) \ =\
x^2_1\,x^2_2\,\Gamma^{\mu}_{S,F}(x_1,x_2) \ee
that do not vanish when $x^2_1=x^2_2=0$\@. 
The simplest solution is
when only $\alpha_5=0$ while others are nonzero, where one gets
\bea 
y^{\mu}_1&\ =\ & \frac{\alpha_1}{\alpha_1+\alpha_2}\,x^{\mu}_1\
, \label{soft1l1} \\
y^{\mu}_2\ &\ =\ &\frac{\alpha_4}{\alpha_3+\alpha_4}\,x^{\mu}_2\
, \label{soft1l2} 
\eea
such that $y^{\mu}_1$ becomes `parallel' to $x^{\mu}_1$ while
$y^{\mu}_2$ is parallel to $x^{\mu}_2$\@.  
The neccessary condition for a singularity, $D^{(1)}=0$, is then 
satisfied only if $x^2_1=x^2_2=0$\@. This solution can be interpreted as a
``soft'' gauge particle, say a gluon, propagating over a finite
(invariant) distance between $y_1$ and $y_2$, such that the
directions of the external particles have not changed after the
emission or absorbtion of the gluon. In momentum space, lines with
negligible momenta are called soft, while in coordinate space soft
lines are those which connect two points at a finite (invariant)
distance. In this case, the scalar/fermion lines are on the lightcone
for a pinch.

If $\alpha_5$ were not equal to zero, the general solution to
Eqs.~(\ref{lan11}) and (\ref{lan12}) is such that
$y^{\mu}_1$, and similarly $y^{\mu}_2$, are given as linear
combinations of $x^{\mu}_1$ and $x^{\mu}_2$\@. Assuming no other
$\alpha_j$ vanishes either, $D^{(1)}=0$ requires then not only $x^2_i=0$ but also
$x^{\mu}_1\cdot x^{\mu}_2=0$\@. These
solutions, however, imply that $x_1$ and $x_2$ are lightlike
separated. Likewise, when only $\alpha_2=0$ or $\alpha_3=0$, the condition
$D^{(1)}=0$ is only satisfied with $x_1$ and $x_2$ lightlike
separated. On the other hand, if $\alpha_1=0$ the solution to
Eq.~(\ref{lan11}) is such that $y^{\mu}_1\propto y^{\mu}_2$ and
by~Eq.~(\ref{lan12})
both $y_1$ and $y_2$ are then parallel to $x_2$ with $D^{(1)}=0$ being
satisfied for $x^2_2=0$\@. However, now the external line~$(x_1-y_1)^2$
can not be on the lightcone, so that this solution does not correspond to
a leading power singularity of the vertex. Similarly, when $\alpha_4=0$,
$y_1$ and $y_2$ will be parallel to $x_1$, and $D^{(1)}=0$ is
satisfied for $x^2_1=0$, so that the external line~$(x_2-y_2)^2$ can not be
on the lightcone either. The solutions with two or more $\alpha_j$
vanishing are either ruled out because of the reasons given above or
because they are equivalent to singularities from zero
lines with~$z^{\mu}_j(y_i)=0$, which we now consider below.

Let us first consider the UV-type singularity of the internal line
connecting the vertex at $y_1$ to the origin,  
$y^{\mu}_1=0$, and look at the conditions for
pinches in the remaining integral over $y^{\mu}_2$\@. Eq.~(\ref{lan12}) is now
satisfied for
\be
y^{\mu}_2\ \ =\ \frac{\alpha_4}{\alpha_3+\alpha_4+\alpha_5}\,x^{\mu}_2\ . 
\label{coly2}\ee
Similarly, when $y^{\mu}_2=0$, Eq.~(\ref{lan11}) gives
\be
y^{\mu}_1\ \ =\ \frac{\alpha_1}{\alpha_1+\alpha_2+\alpha_5}\,x^{\mu}_1\ . 
\label{coly1}\ee
These solutions satisfy the condition $D^{(1)}=0$ if $x^2_2=0$ and 
$x^2_1=0$, separately. According to the
physical interpretation of pinch singularities given in the previous
section, these solutions give ``collinear'' gauge particles that propagate on
the lightcone parallel to one of the external scalars/fermions. We will refer
to such lighlike lines with finite energies and momenta in
the physical interpretation as jet lines.

Next, consider the case when both internal vertices move to the
origin, $y^{\mu}_1=y^{\mu}_2=0$, which makes three lines become zero lines
simultaneously. Again, the vanishing of $D^{(1)}$ for a singularity
requires $x^2_1=x^2_2=0$ with all lines on the lightcone, otherwise
$\alpha_1=\alpha_4=0$\@. 
This solution
represents an ultraviolet (short-distance) divergence, and we call it a ``hard''
solution by analogy to hard scattering.

Among the remaining cases with zero vectors, first consider
$y^{\mu}_2=x^{\mu}_2$ and $y^{\mu}_1=x^{\mu}_1$\@. The solution to
Eq.~(\ref{lan11}) for the former, and to Eq.~(\ref{lan12}) for
the latter, is such that
$y^{\mu}_1$, and similarly $y^{\mu}_2$, is given as a linear
combination of $x^{\mu}_1$ and $x^{\mu}_2$, so that $D^{(1)}=0$ is not
satisfied for these (unless $\alpha_5=0$) because $x^{\mu}_1\cdot
x^{\mu}_2 \neq 0$\@. For $y^{\mu}_1=y^{\mu}_2=y^{\mu}$, both of the
external lines can not be on the lightcone simultaneously. Among the
cases when two propagators have zero arguments at
the same time, $y^{\mu}_1=x^{\mu}_1$ together with $y^{\mu}_2=0$, and 
$y^{\mu}_1=0$ together with $y^{\mu}_2=x^{\mu}_2$ are limiting cases
of collinear solutions in (\ref{coly1}) and (\ref{coly2}),
respectively; while any other combination is ruled out because they
require $x^2_i=0$ and  $x_1\cdot x_2=0$\@. The only possible solution
with three zero lines is the hard solution we found, and there can not be any
other with more zero lines, because $x^{\mu}_1\neq x^{\mu}_2\neq
0$\@. We have finished the list of solutions to Landau conditions for
the leading power singularities of the vertex functions at one loop.

From the solutions to Landau conditions at one loop, one can draw the
conclusion that the divergences of the vertex functions in coordinate
space come from configurations where, whether the gluons are soft or collinear,
the external particles move along {\it rigid, classical} trajectories along the
directions of external points that are located on the lightcone.  
For the singular configurations at higher orders, we will not need 
to solve the Landau equations explicitly. 
Instead, we will make use of the physical interpretation of
the necessary conditions for a pinch singularity given in the previous
section, and confirmed above for the one-loop case.

In the case of the pinch singularities of the vertex
function, to identify an arbitrary pinch surface, we can
use the necessary condition~(\ref{landau2}) that the lightlike lines
of the
corresponding diagram must describe a physical process, where
the two external lines start from the same point, say the origin, 
moving in different directions, toward $x^{\mu}_1$ and $x^{\mu}_2$\@. For the
sake of the argument, suppose the external particles are fermions. Any
gauge field lines that connect them by vertices at finite
distances have to be soft, 
because they cannot be parallel to both. They may still have a hard
interaction at the origin reflecting a
short-distance singularity. The integrals
over the positions of the fermion-gluon vertices will be pinched
either when the gluon and the fermion lines get mutually collinear, or
when the two collinear fermions are connected by the emission of a soft gluon  
as described in the example given at the end
of Sec.~\ref{analysis}\@. 
Likewise, the integrations over the
positions of vertices, to which these collinear gluons are connected,
will be pinched if the other lines connected to these
vertices also become parallel to them, such that all collinear
lines make up a ``jet'' moving in the direction of the external
fermions. The Landau conditions allow these collinear gluons to emit
soft gluon lines that can connect to the other jet as well. Therefore,
the two jets can have hard interactions at very short distances, and
they can only interact by the exchange of soft partons at long
distances at later times.
Eventually, the fermion lines end at the external points
$x^{\mu}_i$, which have to be on the lightcone, $x^2_i=0$, in order that
$D^{(n)}=0$ is satisfied.

To sum up, the
pinch singularities of the integrals for vertex functions in
coordinate space 
come from
configurations where the (time-ordered) vertices, at which either 
soft or collinear gluons are emitted or absorbed, are aligned along
{\it straight} lines going through the `origin' and the external points.
These two lines also
determine the classical trajectories of the external particles in the
Coleman-Norton interpretation. 
The behaviour of the integrals for arbitrary diagrams at higher orders near
the corresponding pinch surfaces will be covered by general power
counting arguments in the next section.

\section{Power counting}\label{power}

In this section, we will apply a power counting technique similar to
the one developed
for momentum space integrals in Ref.~\cite{Sterman:1978bi} to study
the behaviour of the divergences of vertex functions in coordinate
space. We have studied in the previous section the pinch
singularities in the integrals when the
external points are on the lightcone. As the external propagators are
not truncated, even the zeroth order results are
very singular in coordinate space when the external points are on the
lightcone; for instance, the fermionic vertex diverges
as~$1/(x^2_1\,x^2_2)^2$\@.  
Therefore, we will now only consider the external points set
on the lightcone, and look for the degree of divergence of vertex
functions with respect to their lowest-order results. In a sense, we are
looking for any possible divergences in
the residues of the lightcone poles,
by analogy to the residues of single-particle poles in external
momenta of Green functions in momentum space for S-Matrix elements. 
We will show at
the end of this section by power counting arguments that 
vertex functions in coordinate space have at worst 
logarithmic divergences with
respect to their lowest order results at higher orders in perturbation theory.

Since $x^2=0$ does not imply
$x^{\mu}=0$ in Minkowski space, na\"{\i}ve dimensional counting
does not neccessarily bound the
true behaviour of the integrals. As we already mentioned in
Sec.~\ref{analysis}, the divergences of the integrals are related to both
the volume of the space of normal variables and the singularities of
the integrand. Therefore, we will do
the power counting by combining the size of the volume element of
normal variables with that of the integrand, which depends
on both normal and intrinsic variables. In order to estimate the size
of the integrand, we will first approximate the integrals near
the pinch surfaces by keeping in each factor (numerator or
denominator) only the terms of lowest order in normal
coordinates, as their scale goes to zero. Then the resultant integrand
is a homogenous function of normal variables, and
the powers of the normal variables in the homogenous integrals
combined with the {\it normal volume element}  will
give us the bounds on the original integrals.

Suppose $z_1, \dots ,z_n$ and $w_1, \dots ,w_m$ denote the
normal and intrinsic variables for a pinch surface~$S$ of an
integral~$I$\@. The $z_i$ vanish on
$S$ with a scale $\lambda$, while $w_j$ remain finite (as
$\lambda\rightarrow 0$) on
$S$\@. For our discussion for vertex functions, we can choose a single
scale $\lambda$ for properly chosen normal
coordinates in our integrals to do the power counting, although we
should stress that this
does not need to be the case in general. 
The scale~$\lambda$
bounds the size of each normal variable, and measures the ``distance''
of the hypercontour~$\mathcal{H}$ from the pinch surface~$S$\@. 
The homogenous integral $\bar{I}$ near the pinch surface $S$
will have the form\footnote{This form with a delta fuction having the sum of the
squares of the absolute values of the normal variables in its argument
corresponds to bounding the normal space with an $n$-dimensional
sphere with radius $\lambda$\@.} 
\be
\bar{I}  \sim  \int d\lambda^2\int_{\mathcal{H}}\, \left(\prod^n_{i=1}
dz_i\right)\,\delta(\lambda^2-\sum_i|z_i|^2)\int\left(\prod^m_{j=1}dw_j\right)\,\bar{f}(z_i,w_j)
\ , 
\ee
where the homogenous integrand $\bar{f}(z_i,w_j)$ is obtained by
keeping only terms lowest order in $\lambda$ in each factor of the
original integrand such that
\be f(z_i,w_j)\ =\ \lambda^{-d_H}\,\bar{f}(z'_i,w_j)\,\left(1 + 
\mathcal{O}(\lambda)\right) \ , \ee  
for each normal variable~$z_i=\lambda\,z'_i$ with $d_H$ the degree of
homogeneity of $\bar{f}(z_i,w_j)$\@. More
specifically, as we will do the analysis for integrals of the form
of Eq.~(\ref{master1}), $d_H$ equals the sum of the lowest
powers of $\lambda$ in the denominator factors minus that in the numerator
factors in~$f(z_i,w_j)$\@.  
The idea is then to scale out
$\lambda$ from each factor in the homogenous integral, to
count the overall power, and find the behaviour of the integral as
$\lambda\rightarrow 0$, 
\be
\bar{I}  \sim  \int d\lambda\, \lambda^{\gamma -1} \int_{\mathcal{H}}\,\prod^n_{i=1}
dz'_i\,\,\delta(1-\sum_i|z'_i|^2)\int\prod^m_{j=1}dw_j\,\bar{f}(z'_i,w_j)
\ ,
\ee
where the overall degree of
divergence~$\gamma$ is given by
\be \gamma\ =\ n-d_H \ . \label{overalldod} \ee
By overall degree of
divergence, we mean the power of scale parameter~$\lambda$ in the
overall integral when all normal variables have the same scale. 
For $\gamma=0$, the
divergences are logarithmic, while for
$\gamma<0$ there will be power divergences.

If some
set of normal coordinates vanished faster than others on~$S$, say as
$\lambda^2$, that would
only increase the power of $\lambda$ in the volume element of the
normal space, giving a nonleading contribution, unless there are new
pinches in the homogenous integral after dropping terms that are
higher order in the normal variables. These pinches, which could occur
between poles that were
separated by nonleading terms before they were dropped, could enhance
the integrals in principle. However, we will argue that, for our choice
of variables, only pinches of the hard-jet-soft type occur in the
homogenous integrals. 
Notice also that, 
if every normal variable in a
subregion vanishes faster than those of other regions, the power
counting will still be the same for each subregion. 
Before two lines
can produce a pinch singularity, they must approach the hypercontour to the same
scale. Therefore, one can choose a single scale for all
normal variables for the power counting for a particular vertex function near a
pinch surface. We shall now apply this power counting technique for the
fermionic vertex given by Eq.~(\ref{fvertexfnc}), which we will refer
to as the vertex function in the following. As we shall see, power
counting for the scalar vertex is essentially equivalent.

%%%%%%%%%%%%%%%%%%%%%%%%%
\begin{figure}[t]
\centering
\includegraphics[scale=1]{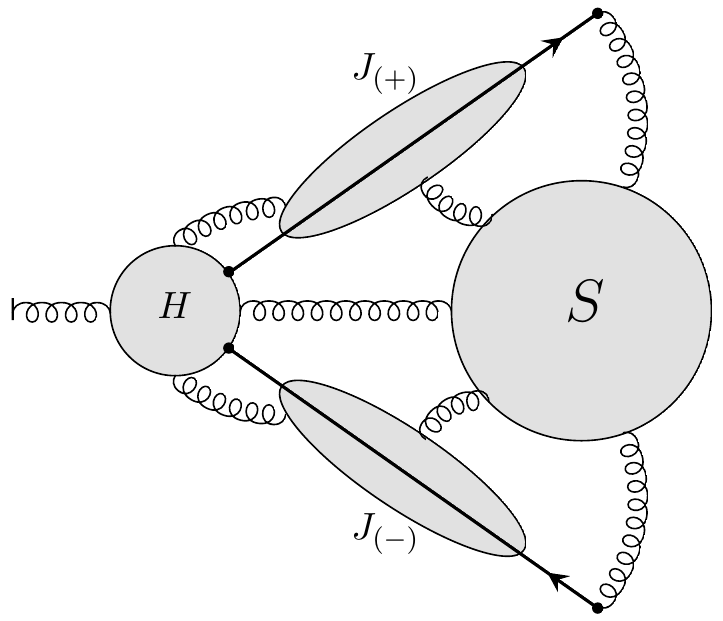}
\caption{Illustration of the soft ($S$), hard ($H$), and jet
  ($J_{(\pm)}$) regions.}
\label{fig:hardjetsoft}
\end{figure}
%%%%%%%%%%%%%%%%%%%%%%%%%%

As we are dealing with lightlike lines, we will use lightcone
coordinates, because the Landau equations for lightcone singularities
can be solved more simply in these coordinates. By a
combination of rotations and boosts, one can put one of the
external points at $x^{\mu}_2=X_2\,\delta^{\mu+}$ and the other one 
at $x^{\mu}_1=X_1\,\delta^{\mu-}$ giving $x^2_1=x^2_2=0$ with 
$x_1\cdot x_2\neq 0$\@. This will set our coordinate system.

For a particular pinch surface~$S$, {\it i.e.} a particular solution to
Landau equations, one has to identify the intrinsic and normal
variables that parametrize the surface and its normal space,
respectively. 
For instance, let us take the solution with a soft gluon
at one loop from Sec.~\ref{form}\@. There, $y^+_2$ and $y^-_1$ are the
intrinsic variables that remain finite, while $y^-_2$, $y^+_1$,
$y^2_{2,\perp}$, and $y^2_{1,\perp}$ are the normal variables that
vanish on the pinch surface. Such lines connecting
vertices at finite distances are part of the soft function. On the other
hand, for the hard solution at one loop, all components of $y^{\mu}_1$ and
$y^{\mu}_2$ were vanishing, and one has only normal variables for
this solution. We will refer to the set of lines connecting a set of
vertices all of whose coordinates vanish as the hard function. Lastly,
in the one loop example, there
were collinear lines. At one loop, for a line in the plus direction,
$y^+_{2}$ is the intrinsic variable while the rest of the
components are normal variables. The set of collinear gluons
together with the external lines to which they are parallel define the
jet function and its direction. 
Note that the limits of
integration of $y^+_2$ here goes from zero up to $X_2$ on this pinch
surface. For $y^+_2>X_2$, the Landau equations can not be
solved. Indeed, $y^+_2>X_2$ does not correspond to a
physical process where all lines move forward in time. In general, in a jet the
limits of integration over the intrinsic variables for a given pinch
surface are not unbounded, but are set according
to the time-ordering of vertices along the jet direction.

In homogenous integrals, the
denominators of the jet lines will be linear, and those of the hard
lines quadratic in normal variables; while the soft lines are of
zeroth order. The lines connecting the hard function
with the jets are linear in normal variables, and we thus count them as
part of the jets. Any line connecting the jets or the
hard part to the soft function is zeroth order in normal
variables, and hence they are counted in the soft function.

The only approximation in writing the homogenous integrand is dropping
terms that are higher order in the normal variables. In fact, such
terms occur only in lines
connecting two different subdiagrams, namely, in lines connecting the jets
to the hard and soft function or those connecting the hard
and soft functions. In order for
our power counting arguments to be valid, there must not be new pinches 
introduced in the homogenous integrals  because of this
approximation. In other words, the Landau
conditions for the homogenous integral to be pinched must have the
same solutions as those of the original integral (up to trivial shifts
or rescalings in some of the variables). We mean by the same
solutions, that the pinch singularities of both have the same physical
picture. Note that soft lines are never pinched in coordinate space,
and thus we
only need to consider the pinches of the lines connecting a jet to the
hard part in the original and homogenous integrals. To this end, one
may consider, for instance, the following integral over
two jet lines, which is a part of a jet function,
\be I=\int d^4y\,\frac{1}{-(x-y)^2+i\ep}\,\frac{1}{-(y-z)^2+i\ep} \ , \label{jetexint} \ee
where $x^{\mu}$, $y^{\mu}$ are jet vertices and $z^{\mu}$ is a hard
vertex\footnote{Here, we have omitted the derivatives at each vertex in
  order to write a simple integrand to illustrate the idea.}\@. 
The Landau conditions for a
pinch from these two lines after Feynman parametrization are given by
\be -\alpha_1\,(x-y)^{\mu}\ +\ \alpha_2\,(y-z)^{\mu}\ =\ 0 \ , \label{orilc}\ee
with $(x-y)^2=(y-z)^2=0$. The only solution to these conditions 
is that all three
vertices are aligned along the jet direction, say the plus direction, and are
parallel to each other. The vertex $z^{\mu}$ is allowed to be hard by
these conditions, {\it i.e.} $z^+$ can also vanish in
Eq.~(\ref{orilc}) at the same rate as the other components. If one
approximates the integral in~(\ref{jetexint}) by a homogenous
integrand with $(y-z)^2\sim
y^2-2y^+z^-$, the condition for a pinch in $y^-$-integral is then the same
as the condition (\ref{orilc}) for the original integral with
$z^+=0$\@. 
The integrals over the transverse components of $y^{\mu}$
can only be pinched at $y^2_{\perp}=x^{2}_{\perp}=0$\@. 
These pinch singularities 
are present in both original and
homogenous integrals. They show up as end-point singularities after
the change of variables with $y^1,y^2\rightarrow y^2_{\perp},\phi$,
which can always be carried out. In general, the pinches of the
homogenous integral correspond to pinches of the original integral
with some of the variables moved to their end-points.

This approximation can fail if $z^+$ becomes comparable to $y^+$ and
$x^+$, or if $y^+$ diminishes like $z^+$, which are actually different
solutions to the Landau conditions corresponding to different pinch
surfaces. Nevertheless, we will see that the result of power counting
will not differ in either case, whether the vertex
$z^{\mu}$ is taken as part of the jet, or included in the
hard part. 
In our analysis, we identify the normal
variables of a pinch surface, group each vertex in a 
certain subdiagram depending on the size of the components of its
position, and do the power counting for the divergence on that
particular pinch surface. 
Generally
speaking, the approximations in the homogenous integrals can change if
two different regions overlap when some vertices escape to a different
subdiagram. However, we will show that the changes in the powers of
the factors of two subdiagrams due to removal of a vertex from one
subdiagram and its inclusion to the other subdiagram cancel each other,
leading to the same conclusion for the overall degree of divergence.

\subsection{Power counting for a single jet}\label{pwcjet}

\begin{figure}[b]
\centering
\includegraphics[scale=1]{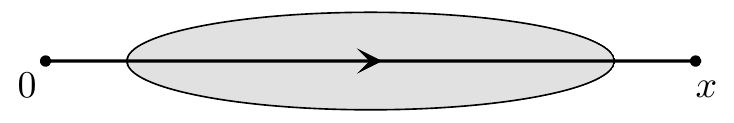}
\caption{A single jet with the topology of a self-energy diagram.}
\label{fig:singlejet}
\end{figure}

Before we do the power counting for the full vertex function in coordinate
space, we begin with the power counting for a single jet, 
with the topology of a self-energy diagram, as depicted
in~Fig.~\ref{fig:singlejet}\@. Let us take a ``dressed'' ultrarelativistic
fermion moving in the plus direction,
so that the plus coordinates of all vertices inside the jet are the intrinsic
variables, while their minus coordinates and squares of transverse
positions are the normal variables. With these choices, all normal
variables appear linearly in jet line denominators. 
The condition that
the vertices have to be on the lightcone for pinches leaves us with
three variables for each vertex. The
azimuthal symmetry around the jet axis allows us to choose the
square of the transverse components as one normal variable. 
We will compute the
contributions to the overall degree of divergence of the
jet, $\gamma_J$, from the normal volume element, the denominators, and
the numerators of the jet function as defined in
Eq.~(\ref{overalldod}) for a generic singular integral.

%%%%%%%%%%%%%%%%%%%%%%%%%
%%%%%%%%%%%%%%%%%%%%%%%%%%

For every
integration over the positions of the three- and four-point vertices
inside the jet, one needs to add $+2$ to $\gamma_J$, that is, $+1$ for
each normal variable (transverse square and minus component)\@. 
In $D=4-2\vep$ dimensions, however, the power for
transverse square components is $+(1-\vep)$\@.  
In the homogenous jet function, the denominator of
each gauge field line contributes a term $-(1-\vep)$ while that of each
fermion line contributes $-(2-\vep)$ to $\gamma_J$, since the massless fermion
propagator in coordinate space is given, as in Eq.~(\ref{1loopvrtxF}), by
\be
S_F(x)=\slashed{\partial}\,\Delta_F(x^2)=\frac{\Gamma(2-\vep)}{2\,\pi^{2-\vep}}\,\frac{\slashed{x}}{(-x^2+i\ep)^{2-\vep}}
\ .\ee
We are interested in the degree of divergence with respect to
the lowest order result,
so we will multiply the diagrams of~Fig.~\ref{fig:singlejet} 
by $[\slashed{x}/(-x^2+i\ep)^{2-\vep}]^{-1}$\@.
Equivalently, we add a
term $+(2-\vep)$ to~$\gamma_J$ to cancel the lightcone divergence
of the lowest order diagram, which is simply the fermion propagator.

Now we
consider the numerator suppressions. In order to get the leading
divergences, in the numerators we will keep only the terms lowest order in
normal coordinates, which therefore give the least suppression. To
begin, we note that there
is a factor contributing to the numerator
from each fermion-gluon vertex at a point $y^{\mu}_n$ of the form,
\be\begin{split}
  (\slashed{y}_{n+1}-\slashed{y}_{n})\,\gamma^{\mu}\,&(\slashed{y}_{n}-\slashed{y}_{n-1})\
  =
  \\
& 2\,(y_{n+1}-y_n)^{\mu}\,(\slashed{y}_{n}-\slashed{y}_{n-1}) -
\gamma^{\mu}\,(\slashed{y}_{n+1}-\slashed{y}_{n})\,(\slashed{y}_{n}-\slashed{y}_{n-1})
\ . \end{split}\label{ffgvertex}\ee
Here, the first term is unsuppressed as it is, although the vector
with the index~$\mu$ must form an
invariant with some other vector. 
At the same time, terms proportional to $\gamma^{\mu}$ in Eq.~(\ref{ffgvertex})
either vanish by $(\gamma^{\mp})^2=0$ or
vanish at least as the transverse coordinates of one of the vertices, which
are at the order of~$\lambda^{1/2}$\@. In the case of scalar,
instead of gauge field lines, with Yukawa couplings to fermions, for
each two-fermion-scalar vertex there
would be a factor of
$(\slashed{y}_{n+1}-\slashed{y}_{n})\,(\slashed{y}_{n}-\slashed{y}_{n-1})$
in the numerator, giving the same power-counting suppression of at
least $\lambda^{1/2}$\@.

In addition to numerator factors from
fermion lines, there are factors from three-gluon vertex functions. 
A vertex at~$z^{\mu}_m$ combines with gluon propagators to give terms
that (dropping overall factors) can be written as
\be v_{3g}(z_m,\{y_i\})\ =\
\ep_{ijk}\,g^{\mu_i\mu_j}\,\Delta(z_m-y_i)\,\Delta(z_m-y_j)\,\partial^{\mu_k}_{z_m}\Delta(z_m-y_k)
\ , \ee
where the $y_i$ are the positions of the other ends of the lines. 
Acting on the gluon lines, the derivatives bring vectors from their coordinate
arguments to the numerator, while 
increasing the power of a denominator by one. These vectors also must form
invariants in the numerator, either among themselves or with the Dirac
matrices of the fermion-gluon vertices. Suppose we let $z_i$
denote the position of the $i$th three-gluon-vertex, and $y_j$ the
position of the $j$th fermion-gluon vertex. 
The numerator is then a product of linear combinations of
invariants of the form $\slashed{z}_i$, $\slashed{y}_j$, and $z_i\cdot
z_{i'}$\@. Referring to Eq.~(\ref{ffgvertex}), one could also get factors with
$z_i\cdot y_j$\@. Each such invariant made out of two vectors is linear in
normal variables. 
One can see then that each fermion-gluon vertex
suppresses the numerator by $\lambda^{1/2}$ at least, while every pair of
three-gluon vertices produce an invariant suppressing the numerator by
$\lambda$, while if a three-gluon vertex is contracted with a Dirac
matrix at a fermion-gluon vertex, it also suppresses the numerator by
$\lambda^{1/2}$ at least. 
Hence, the contribution of the numerators to $\gamma_J$ is given by
\be \gamma_{n_J}\ \geq\ \frac{1}{2}(V^f_3+V^g_3)\ =\ \frac{1}{2}V_3 \ ,\ee
with $V^f_{3}$ the number of fermion-gluon vertices and $V^g_3$
the number of three-gluon vertices inside the jet. Adding all
contributions, we obtain a lower bound for the overall degree of
divergence of this fermion jet,
\be \gamma_J\ \geq\
2(V_3+V_4)+2-N_g-2N_f-V^g_3+\frac{1}{2}V_3+\mathcal{O}(\vep)  \
. \label{jetpower}  \ee
Here, $V_3$ ($V_4$) is the total number of three-point (four-point)
vertices, while $N_g$ ($N_f$) is the number of gluon (fermion) lines in
the jet. We can use the graphical identity,
\be 2N\ =\ E+\sum_{i=3,4}i\,V_i \ , \label{graphidentity}\ee
in Eq.~(\ref{jetpower}), which relates the number of lines of a diagram to
the numbers of its various kinds of vertices, where $E$ is the number
of external lines of the jet. A single jet has two
external lines, $E=2$, one connected to the external point $x^{\mu}$ and the
other to the origin. Combining the number of lines $N=N_f+N_g$ and the
number of three-point vertices $V_3=V^f_3+V^g_3$, we can rearrange the terms in
Eq.~(\ref{jetpower}),
\be \gamma_J\ \geq\ \frac{3}{2}V_3 + 2V_4 -N +2 +V^f_3 -N_f \ , \ee
which, using the graphical identity~(\ref{graphidentity}), can be
reduced to
\be \gamma_J\ \geq\ 1+V^f_3-N_f \ . \ee
Here, we note that, because at each fermion-gluon
vertex one fermion line enters and one exists, the number of fermion
lines in the jet are equal to one plus the number of fermion-gluon
vertices in the jet. Therefore, our power counting results in 
\be \gamma_J \ \geq\ 0 \ . \label{gammajet}\ee
Thus, a fermion jet in coordinate space with the topology of a self-energy
diagram can have at worst logarithmic divergence. In contrast to the
power counting in momentum space, we did not count the number of loops
nor did need to use the Euler identity. Note that the
power counting for a scalar jet gives the same result, because the
derivatives at two-scalar-gluon vertices in a scalar jet
correspond to the derivatives from fermion propagators in a
fermion jet. Similarly, the two-scalar-two-gluon ``seagull'' couplings
have no numerator factors, and are counted like the four-point gluon
couplings.\footnote{By the
arguments given in the main text, the overall degree of divergence of
a scalar jet is bounded from below by
\[\gamma_{SJ}\ \geq\
2(V_3+V_4+V^{2g2s}_4)+1-N_s-N_g-V^g_3-V^{sg}_3+\frac{1}{2}(V^g_3+V^{sg}_3)+\mathcal{O}(\vep)
\ , \]
with $N_s$ the number of scalar lines, $V^{2g2s}_4$ the number of
two-scalar-two-gluon vertices, and $V^{sg}_3$ the number of
two-scalar-gluon vertices. Each term on the right hand side cancels by
Eq.~(\ref{graphidentity}) such that $\gamma_{SJ}\geq 0$ in this case as well. 
}

If we had kept the
terms at $\mathcal{O}(\vep)$ in the power counting of
Eq.~(\ref{jetpower}), we would have derived a bound
\be \gamma_J\geq (\frac{1}{2}V_3+V_4)\vep \ , \ee 
which shows that these collinear
singularities are regulated also by $\vep > 0$ in coordinate space. 
No IR regularization is neccessary after UV renormalization when
the external points are taken to the lightcone.
However, in the Fourier transform of the vertex function
for S-Matrix elements in momentum space, the divergences in $p^2=0$
will require IR regularization with $\vep < 0$ when the
external points are integrated to infinity.

The power counting above and the result~(\ref{gammajet}) hold in the presence of
self-energies inside the jet as well. For instance, cutting a gluon
line in the jet and inserting a fermion loop does not change
$\gamma_J$, because the changes due
to extra fermion denominators are canceled by the terms
for integrations over the
positions of these two new vertices, while the denominator of the
extra gluon 
cancels the contribution of fermion numerators to $\gamma_J$ since
\mbox{$\slashed{y}(-\slashed{y})=-y^2\sim\mathcal{O}(\lambda)$} with
$y^{\mu}$ the difference of the positions of the
two vertices. In the case of inserting a gluon or a ghost loop, a similar
cancellation occurs. The denominators of the two new
lines in the loop each have a lower power by one compared to fermion
lines, but there are now 
two derivatives at the new vertices raising those powers. 
A different power counting is needed for the case when such
self-energies shrink to a point, that is $y^+\rightarrow 0$ in
the example above for a jet in the plus direction. 
When renormalization has
been carried out, such UV-divergences are removed by local counter-terms.

\subsection{Overall power counting for the vertex function}\label{opc}

We are now ready to continue with the overall power counting for the
vertex function including two jets, a soft subdiagram, and a hard
subdiagram as in Fig.~\ref{fig:hardjetsoft}\@. We will do the analysis
for the fermionic vertex function. As in the previous subsection, the
counting is the same for the scalar vertex. It also straightforwardly
extends to any amplitude for wide-angle scattering.

The homogenous soft function is independent of normal
variables, and by dimensional counting it is finite for fixed external
points. 
We introduce the notation $J^H_{(\pm)g}$ and $J^H_{(\pm)f}$ to denote
the numbers of vector and fermion lines, respectively, that connect
the hard subdiagram to the jets in the $\pm$~direction. In these
terms, we also define
\bea 
J^H_{g,f} &=& J^H_{(+)g,f} + J^H_{(-)g,f} \ , \\
J^H &=& J^H_g + J^H_f \ , \label{jhlines}
\eea
where $J^H$ is the total number of lines attaching both jets to the
hard subdiagram. Similarly, we define for the lines connecting the
jets to the soft subdiagram, 
\bea 
S^J_{g,f} &=& S^J_{(+)g,f} + S^J_{(-)g,f} \ , \\
S^J &=& S^J_g + S^J_f \ , \label{sjlines}
\eea
and lastly for the lines connecting the soft and hard subdiagrams, 
\be S^H\ =\ S^H_g + S^H_f  \ . \ee 
Recall that all
components of the vertices in the hard
function vanish together, so that hard lines are quadratic in normal
variables. Similar to Eq.~(\ref{jetpower}) for a single jet, 
the overall degree of divergence for the vertex function relative to
the lowest order diagram can be written as 
\be\begin{split}
\gamma_{\Gamma} \ \geq & \ 
4(V^H_3+V^H_4)-2N^H_g-3N^H_f-V_{3g}^H  \\
     &  +\sum_{i=+,-} \Big[
     2(V^{J_{(i)}}_3+V^{J_{(i)}}_4)+2-N^{J_{(i)}}_g-2N^{J_{(i)}}_f-V^{J_{(i)}}_{3g}+n^{J_{(i)}}\Big]+\mathcal{O}(\vep)
     \ , 
\end{split}\label{gammaoverall}\ee
where $n^{J_{(\pm)}}$ denotes the numerator contributions from the
jet in the $\pm$~direction. The terms labeled $H$ are contributions from the
hard part, followed by contributions from the two jets. Note that there
are no contributions from integrations over the positions of soft
vertices here, because all of their components are intrinsic variables.

In the hard part, every three-gluon vertex produces a vector that
must be proportional to a linear combination of the position vectors,
$z_i^{\mu}$ of vertices in the hard
subdiagram. These are all normal variables, and are hence
order~$\lambda$. These vectors may form invariants with a jet or a
soft vertex suppressing the numerator by
$\lambda$, or two of them may form an invariant at
$\mathcal{O}(\lambda^2)$\@. Thus, each hard three-gluon vertex
contributes $+1$ to scaling of the numerator while their derivatives
increase the power of a gluon denominator that is quadratic in $\lambda$\@. In
total, they contribute $-V^H_{3g}$ to $\gamma_{\Gamma}$\@.

The numerator contributions of jets are somewhat 
different compared to Eq.~(\ref{jetpower}), because the vectors
arising from the derivatives of three-gluon vertices inside the jets can now 
form invariants with vectors from three-gluon vertices in the hard or
soft part, or from the opposite moving jet. At lowest order in normal
variables, the invariants resulting from contracting a jet vertex with
a soft vertex are zeroth order in normal variables, 
while those from a jet and a hard vertex are linear, 
which, however, we have already counted
in~(\ref{gammaoverall}) among the contributions from the hard
part. There can be at most
$J^H_g$ such vectors to form out-of-jet invariants, as
many as the number of lines connecting jets to the hard part. The
polarization of any of the $S^{J}_{g}$ soft gluons connecting the
jets to the soft function does not produce an invariant that 
contributes to $n^J=n^{J^{(+)}}+n^{J^{(-)}}$, and the
fermion-gluon vertices in the jets where a soft fermion line attaches
do not always give a suppression in the numerator. For the minimum
numerator suppressions, we can
thus subtract $J^H_{g}+S^J_g+S^J_f$ from the total number of three-point
vertices in $n^J$, 
\be n^{J}\ \geq \
\frac{1}{2}(V^{J}_3-J^H_{g}-S^{J})  \ , \label{numjets}
\ee
where we use the notation of Eqs.~(\ref{jhlines}) and
(\ref{sjlines})\@.

We can again apply the graphical identity in~(\ref{graphidentity}) to
the terms in Eq.~(\ref{gammaoverall}) for the jets and the hard
subdiagram separately. The $E^H$ external lines of the
hard function are either jet or soft lines,
\be E^H = S^H+J^H \ , \ee
where we assume for this discussion that the minimum of the fermion
lines connecting the jets to the hard part is two, $J^H_f\geq 2$, one
from each jet. Pinch surfaces where only gluons attach the hard part
to the jets in the reduced diagram are also possible, and may be
treated similarly, with equivalent results. These external lines
must be added to the number of hard lines, $N^H=N^H_{g}+N^H_{f}$, in
the identity for the total number of lines connected to the hard part, 
\be
 2N^H+E^H \ = \ 2V^H_2+3V_3^H+4V_4^H \ . \label{hardnumber}
\ee
Here, we consider the vertex of the external current as a
two-point-vertex, so that $V^H_2=1$\@. The total number of jet lines
are related to the number of vertices in both jets by
\be 2N^J+S^J\ =\ 2+J^H+3V^J_3+4V^J_4 \ , \label{jetnumber}\ee
where the number of (soft) lines, $S^J$, connecting the jets to the
soft part is added to the number of jet lines. Removing the
contributions from the gluon lines and vertices they attach in 
Eq.~(\ref{hardnumber}), one can find a relation between the number of
fermion vertices and the number of fermion lines in the hard subdiagram,
\be V^H_{3f}\ = \ N^H_f +\frac{1}{2}(S^H_f+J^H_f-2) , \label{hardfvrtx}\ee
while a similar relation can be found from (\ref{jetnumber}) for jets
\be V^J_{3f}\ =\  N^J_f +\frac{1}{2}(S^J_f-J^H_f-2) \ . \label{jetfvrtx}\ee

To derive a lower bound for $\gamma_{\Gamma}$ in Eq.~(\ref{gammaoverall}) it is
convenient to begin by applying (\ref{hardnumber}) to the `$H$'~terms
of $\gamma_{\Gamma}$, and Eq.~(\ref{jetnumber}) to the jet terms. Then, we can
readily use the relations of fermion lines to vertices,
(\ref{hardfvrtx}) and (\ref{jetfvrtx}), for the hard subdiagram and for
the jets, respectively, and the numerator inequality~(\ref{numjets}),
to derive a lower bound for the overall degree of divergence of the
vertex function, 
\be \gamma_{\Gamma} \ \geq\ S^H_g + \frac{3}{2}S^H_f + \frac{1}{2}(S^J_f +
J^H_f - 2)  \ . \ee
The condition for a
(logarithmic) divergence is then that no line can connect the
hard subdiagram directly to the soft subdiagram and that only a single
fermion attach each jet to the hard subdiagram. This corresponds to
similar results found for pinch surfaces in momentum
space~\cite{Sterman:1978bi,Sterman:1995fz}\@. The soft and hard
subdiagrams can only interact through jets. 
Moreover, when the lower bound is saturated, using the same relations above,
the leftover terms in $\gamma_{\Gamma}$ that are at the order
of~$\vep$ can be shown to be equal to
\be \gamma_{\Gamma}^{\mathcal{O}(\vep)} = \Big(\frac{1}{2}V_3^J+V^H_3+V^J_4+2V^H_4
- \frac{1}{2}[S^J_g+J^H_g]\Big)\vep \ . \ee
For each line connecting the soft part to a jet there is a vertex in
the jet, while for each line connecting a jet to the hard part there is a
hard vertex. Thus, there will be enough vertices left
over to make the coefficient of $\vep$ positive. Therefore, the
logarithmic divergence of the vertex function is regulated by $\vep>0$
in coordinate space.

\begin{figure}[t]
\centering
\includegraphics[scale=1]{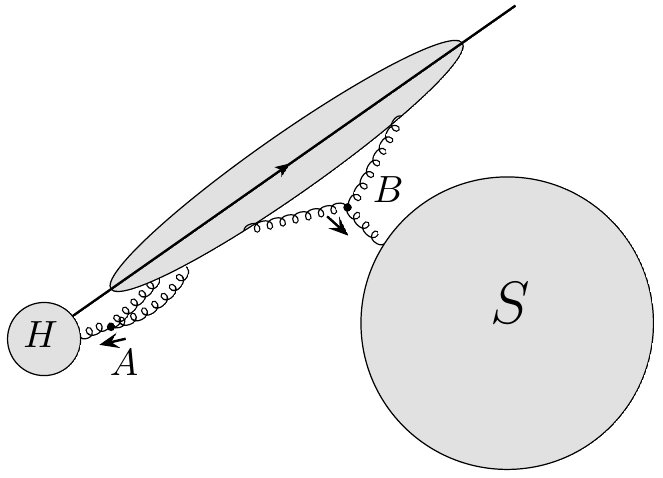}
\caption{A jet vertex $A$ gets captured by the hard part, while
  another vertex $B$
  escapes to the soft part. Power counting for this pinch surface
  gives the same result as when the vertices $A$ and $B$ were inside
  the jet. The other jet in the opposite direction is suppressed here.}
\label{fig:vri}
\end{figure}

We shall now consider the changes in the power counting due to removal
of a vertex from one subdiagram and its inclusion to the other
subdiagram. This will happen at the boundary of integration in
intrinsic variables. Suppose that one of the jet vertices connected to the hard
part gets captured by the hard part and becomes part of it, or a vertex
in the jet escapes to the soft part, as depicted in
Fig.~\ref{fig:vri}\@. In the first case, the line, which
used to connect the hard part to one of the jets has become a hard
line, while the other jet lines attached to that vertex now connect the
hard part and the jet. Thus, if the vertex that gets captured is a
three-gluon vertex, $N^H_g$, $V^H_{3g}$, and $J^H$ each 
change by $+1$, while $N^J_g$ and $V^J_{3g}$ change by
$-1$\@. Likewise, if a four-point vertex gets captured, $N^H_g$ and $V^H_4$ 
increase by $+1$ but $J^H$ now increases by $+2$, while $N^J_g$ and
$V^J_4$ change by $-1$\@. These
changes, however, cancel exactly in Eq.~(\ref{gammaoverall}) using the bound
for the jet numerator contributions $n^J$ in
Eqs.~(\ref{numjets}) for the most divergent
configurations. Similarly, if a three-gluon vertex in the jet escapes
to the soft part, it pulls two lines out of the jet making them soft,
hence $N^J_g$ and $V^J_{3g}$ change by $-2$ and $-1$, respectively, while $S^J$
increases by $+1$\@. For a four-point vertex that escapes the jet and
joins the soft part, $N^J_g$ and $V^J_4$ decrease by $-3$ and $-1$,
respectively, while $S^J$ increases by $+2$\@. These
changes also cancel in Eq.~(\ref{gammaoverall}) for the most divergent
configurations. Note also that when a hard vertex escapes to the soft
part the leftover changes in~Eq.~(\ref{gammaoverall}) are equal to
$\Delta S_H$, the change in the number of lines connecting the hard
and soft parts. Therefore, the leading behaviour does not change, even
if two different subdiagrams do overlap, as was asserted at the
beginning of this section.

To conclude, we have shown
by power counting arguments that the vertex function in coordinate
space can diverge at worst logarithmically times overall lowest-order
behavior. This logarithmic divergence requires $D<4$ in dimensional
regularization.

\section{Approximations and Factorization}

A fundamental consequence of the structure of pinch surfaces is the
factorization of soft gluons from jets and jet gluons from the hard
part. This is shown in momentum space by the use of Ward
identities~\cite{Collins:1989gx,sterman-qft,collinspQCD}\@. In this
section, we show how the same Ward identities, as they appear in
coordinate space, result in the factorization of soft, jet and hard
functions.

\subsection{Hard-collinear approximation}

Having identified the jet and hard regions
that can give divergences in coordinate-space integrals in the
previous section, we now construct a coordinate space {\it
  ``hard-collinear''} approximation to the integral, which enables
factorization of the jet and hard functions at  
the leading singularities. Recall that the only
approximation made for writing a homogenous integrand to do the
power counting for these two regions was dropping the terms higher
order in normal variables in lines connecting the jets to the
hard part. Thus, the approximation one needs is made on
the propagators of these jet lines attached to the hard part. We shall
explain this {\it ``hard-collinear''} approximation with the example
of the following integral,
\be I(y)\ =\ \int
d^4z\,J^{\nu}(y)\,g_{\nu\rho}\,D^{\rho\mu}(y-z)\,H_{\mu}(z) \
,\label{hard-col1} \ee 
where $J^{\nu}$ denotes a jet function with a direction $\beta^{\nu}$,
$D^{\rho\mu}(y-z)$ the propagator of the line that connects a jet
vertex at $y$ to a hard vertex at $z$, and $H_{\mu}(z)$ a hard
function. We raise and lower the indices by the Minkowski
metric. Here, we have suppressed the dependence on
other vertices, which are also integrated over. The integral
in~(\ref{hard-col1}) will have divergences when the jet moves in the plus
or minus lightcone direction and all coordinates of the hard function
vanish. In this limit, we can approximate this integral by picking out
the large component of the jet, by replacing $g_{\nu\rho}\rightarrow
\beta'_{\nu} \beta_{\rho}$ where $\beta^{\mu}=\delta^{\mu+}$ and
$\beta'^{\nu}=\delta^{\nu-}$, 
\be I(y)\ \sim\ \int
d^4z\,J^{\nu}(y)\,\beta'_{\nu}\,\beta_{\rho}\,\bar{D}^{\rho\mu}(y-z)\,H_{\mu}(z)
\ . \label{hard-col2} \ee
In the gluon propagator, $\bar{D}$, we
neglect the smaller terms coming from the hard vertex. Let us take the
jet to be in the plus-direction, then the dependence on $z^{\mu}$ in the
argument of the propagator will be largely through $z^-$, the component of
the hard vertex in the opposite direction, because 
\be (y-z)^2 =  2y^+(y^--z^-)-y^2_{\perp}+\mathcal{O}(\lambda^{3/2}) \ , \ee
for $y^+\gg z^+$ and $y^2_{\perp}\gg z^2_{\perp}$\@. We then write the
propagator as
\bea
\beta_{\rho}\, D^{\rho\mu}(y-z)& = &D^{-+} (y-z) \beta^{\mu} \ , 
\nonumber\\
\bar{D}^{-+}(y-z)&=&\frac{\partial}{\partial
  z^-}\int^{z^-}_{\infty}d\sigma\,D^{-+}\big(  2y^+(y^--\sigma\,\beta'^-)-y^2_{\perp}    \big)
\nonumber \ , \\
 & \equiv & \partial_{z^-}\mathscr{D}(y,z^-) \ . 
\eea
For this representation, we should take $\vep <0$ in $\bar{D}^{-+}$\@. 
One can now integrate by parts in Eq.~(\ref{hard-col2}), so that
$-\partial_{z^-}$ acts on the hard function~$H_{\mu}(z)$,
\be I(y)\ \sim\ \int
d^4z\,J^{+}(y)\,\mathscr{D}(y,z^-)\big(-\partial_{z^-}H^-(z)\big) \ . \ee
There are
no boundary terms as a result of integrating by parts,
because in the hard function $H_{\mu}(z)$ there must be at least one
propagator that vanishes at $z^-=\pm\infty$\@. Furthermore, we can add
to the integrand the derivatives with respect to other components of
$z^{\mu}$ such that we now have a full gradient~$\partial^{\mu}_z$
acting on the hard function~$H_{\mu}(z)$. Because the jet function and
$\mathscr{D}(y,z^-)$ 
do not depend on $z^+$ and $z_{\perp}$, these added terms are total
derivatives and vanish after the integration. The result of our
approximation can then be expressed by 
\be I(y)\ \sim\ \int
d^4z\,\big(J^{\nu}(y)\,\beta'_{\nu}\big)\,\mathscr{D}(y,z)\big(-\partial^{\mu}H_{\mu}(z)\big)
\ . \ee
In other words, we have replaced the propagator of the gluon escaping from
the jet to the hard part by
$D^{\nu\mu}(y-z)\rightarrow\mathscr{D}(y,z)\beta'^{\nu}\partial^{\mu}_z$
with $\beta'^{\nu}$ being a vector in the opposite direction of the
jet. The momentum-space analog of such a gluon is called
``longitudinally'' or ``scalar polarized'', and is
associated with the scalar operator $\partial_{\mu}A^{\mu}(x)$ in coordinate
space.

\begin{figure}[t]
\centering
\includegraphics[scale=1]{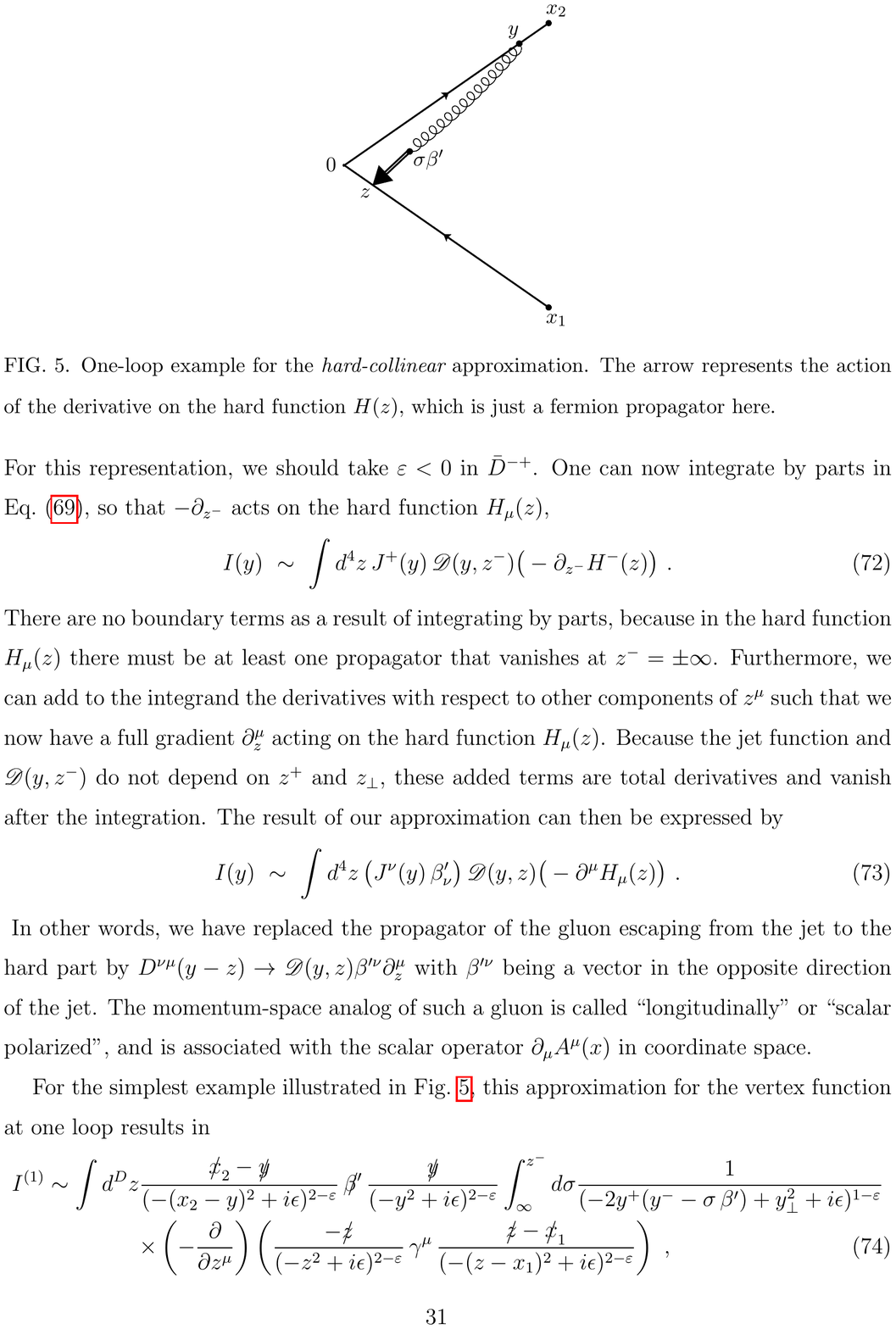}
\caption{One-loop example for the {\it hard-collinear}
  approximation. The arrow represents the action of the derivative on
  the hard function $H(z)$, which is just a fermion propagator here.}
\label{fig:hardcolfig}
\end{figure}

For the simplest example illustrated in~Fig.~\ref{fig:hardcolfig}, 
this approximation for the vertex function at one loop results in
\be\begin{split}
I^{(1)}\sim\int
d^{D}z&
\frac{\slashed{x}_{2}-\slashed{y}}{(-(x_2-y)^2+i\ep)^{2-\vep}}\,\slashed{\beta'}\,\frac{\slashed{y}}{(-y^2+i\ep)^{2-\vep}} \\
 & \times\int^{z^-}_{\infty}d\sigma\frac{1}{(-2y^+(y^--\sigma\,\beta')+y^2_{\perp}+i\ep)^{1-\vep}}\\
 & \quad\times\left(-\frac{\partial}{\partial
    z^{\mu}}\right)\left(
\frac{-\slashed{z}}{(-z^2+i\ep)^{2-\vep}}\,\gamma^{\mu}\,\frac{\slashed{z}-\slashed{x}_{1}}{(-(z-x_1)^2+i\ep)^{2-\vep}}\right)
\ ,
\end{split}\ee
where we have omitted the incoming current, integrations over jet vertices,
and numerical factors.
After acting with $\partial^z_{\mu}$, there are two terms with a
relative sign coming from the action of the derivative on either
fermion propagator, canceling them in turn by the massless Dirac equation
\mbox{$\slashed{\partial}S_F(x)=-\delta^D(x)$}, 
\be\begin{split}
I^{(1)}\sim\int &
d^{D}z\,\frac{\slashed{x}_{2}-\slashed{y}}{(-(x_2-y)^2+i\ep)^{2-\vep}}\,\slashed{\beta'}\,\frac{\slashed{y}}{(-y^2+i\ep)^{2-\vep}}\\
 &\times\int^{z^-}_{\infty}d\sigma\frac{1}{(-2y^+(y^--\sigma\,\beta')+y^2_{\perp}+i\ep)^{1-\vep}} \\
& \quad\times\left(\frac{-\slashed{z}}{(-z^2+i\ep)^{2-\vep}}\,\delta^D(z-x_1)\,-\,
\delta^D(z)\,\frac{\slashed{z}-\slashed{x}_{1}}{(-(z-x_1)^2+i\ep)^{2-\vep}}\right)
\ . 
\end{split}\ee
After integrating
over $z$, the location of the attachment of the ``scalar polarized''
gluon, using the delta functions, the two terms differ only in the
upper limits of the $\sigma$ integrals, which can be combined so that
the the remaining leading term is given~by
\be\begin{split}
I^{(1)}\ \sim\  -
\frac{\slashed{x}_{2}-\slashed{y}}{(-(x_2-y)^2+i\ep)^{2-\vep}}&\slashed{\beta'}\frac{\slashed{y}}{(-y^2+i\ep)^{2-\vep}}\frac{-\slashed{x}_{1}}{(-x_1^2+i\ep)^{2-\vep}}
\\
&\hspace{-1.5cm}\times \left(\int_{0}^{x^-_1}d\sigma\frac{1}{(-2y^+(y^--\sigma\,\beta')+y^2_{\perp}+i\ep)^{1-\vep}}\right) \ . \end{split}\ee
Therefore, after the {\it hard-collinear} approximation the ``scalar
polarized'' gluon has been factored onto an eikonal line in the
opposite direction from the jet of which it is a part such that
the jet is now factorized from the rest of the diagram. Note that the
integration over the eikonal line is a scaleless integral, which in the limit
$x^-_1\rightarrow\infty$ will be defined by its ultraviolet pole only.

The {\it hard-collinear} approximation also allows us to apply the
basic Ward identities of gauge theories directly to the leading singularity of
the vertex function in order to factor the ``scalar polarized''
gluons from the hard function. This reasoning was used in the proofs
of factorization in gauge theories in momentum
space~\cite{Bodwin:1984hc,Collins:1985ue,Collins:1989gx} 
and the same reasoning applies here. The
Ward identity that we need is given by
\be \langle\mathrm{out}|T\Big(\partial_{\mu_1}A^{\mu_1}(x_1)\,\cdots\, \partial_{\mu_n}A^{\mu_n}(x_n)\Big)|\mathrm{in}\rangle\ =\ 0  \ , \label{ward1}\ee
where $|\mathrm{in}\rangle$ and $\langle\mathrm{out}|$ are physical
states involving particles of fermion and gauge fields with physical
polarizations only. The gauge field
$A^{\mu}(x)$ can be abelian or non-abelian, the above matrix element
relation involving scalar polarized gauge fields holds at each order
in perturbation theory after the sum over all contributing
diagrams~\cite{'tHooft:1971fh,'tHooft:1972ue}\@.

\begin{figure}[b]
\centering
\includegraphics[scale=1]{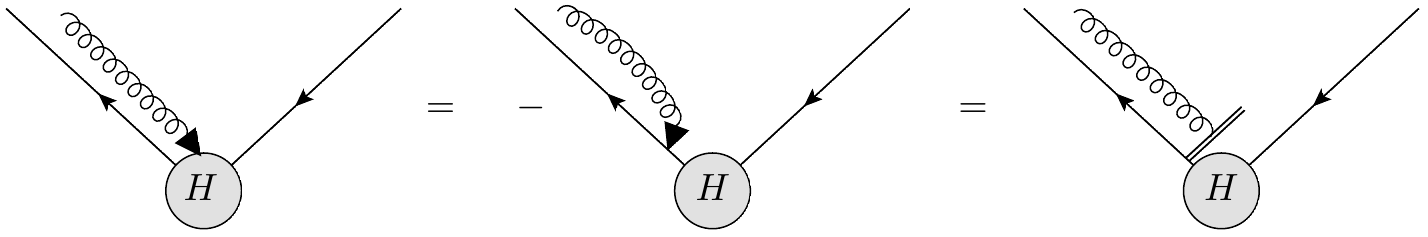}
\caption{Factorization of one ``scalar polarized'' gluon from the hard
subdiagram.}
\label{fig:hjfactor}
\end{figure}

At higher orders, the external lines of the hard function will be two
physical fermion lines on the lightcone, one entering and the other
exiting the hard function, and some number of ``scalar polarized'' gauge field
lines with derivatives acting on the hard function. Consider the
case with one such gluon line connected to the hard part as in
Fig.~\ref{fig:hjfactor}\@. 
This diagram
is equal to minus the diagram where the scalar-polarized gluon is
acting on the fermion line in the same direction by the Ward
identity~[Eq.~(\ref{ward1})], which gives a factored gluon onto an eikonal
line as we showed above at lowest order. This summarizes the argument,
which was also extended to
arbitrary number of gluon lines, for the proof of factorization of
jets (collinear singularities) from the hard function in momentum
space, see the review of Ref.~\cite{Collins:1989gx}\@. The same
factorization can be shown in coordinate space after the {\it
  hard-collinear} approximation described above using the Ward
identity~[Eq.~(\ref{ward1})], in the same way as in momentum
space.

\subsection{Soft-collinear approximation}

One may also do a {\it ``soft-collinear''} approximation for the lines that
connect the jets to the soft function such that the collinear singularities of
jets will be factored from the finite soft function. We will follow
the same reasoning and repeat the same steps as we did for the
``hard-collinear'' approximation. To avoid repetition, we will skip
those details of our arguments that were explained in the previous
section.
In analogy to the integral in
(\ref{hard-col1}), now consider
\be I(z)\ =\ \int
d^4y\,S_{\mu}(z)\,D^{\mu\nu}(z-y)\,g_{\nu\rho}\,J^{\rho}(y) \ .\label{soft-col1} \ee
Here, $S_{\mu}(z)$ denotes the soft function, and $J^{\rho}$ denotes a
jet function with a direction $\beta^{\rho}$\@. $D^{\mu\nu}(z-y)$ is
the propagator of the line that connects a soft vertex at $z$ to a vertex in
the jet at $y$, which are at a finite distance from each
other. We suppress the dependence of either function on
other vertices as before. In this integral, the only singularities of
the integrand are contained in the jet function. Again, to approximate
this integral, we will drop
the small terms in the argument of the progator, and pick up only the
large numerator component of the jet,
\be I(z)\ \sim\ \int
d^4y\,S_{\mu}(z)\,\bar{D}^{\mu\nu}(z-y)\,\beta_{\nu}\,\beta'_{\rho}\,J^{\rho}(y)
\ ,\label{soft-col2} \ee
with $\beta^2=\beta'^2=0$ and $\beta\cdot\beta'=1$\@. Suppose the jet is in the
plus-direction, then following our power counting, $z^-\gg y^-$ and
$z^2_{\perp}\gg y^2_{\perp}$ such that 
\be (z-y)^2 = 2(z^+-y^+)z^--z^2_{\perp}+\mathcal{O}(\lambda^{1/2})\ . \ee
Thus, this time we write the propagator connecting
the soft part to a jet as
\bea
 D^{\mu\nu}(z-y)\,\beta_{\nu} & = & \beta^{\mu}\,D^{+-} (y-z) \ , 
\nonumber\\
\bar{D}^{+-}(z-y)& = & \frac{\partial}{\partial
  y^+}\int^{y^+}_{\infty}d\sigma\,D^{+-}( z-\sigma\,\beta)
 \ \equiv \ \partial_{y^+}\mathscr{D}(z,y^+) \ . 
\eea
Using the steps above, we integrate by parts in Eq.~(\ref{soft-col2}),
and then add to the integrand
the derivatives with respect to other components of $y^{\mu}$ as
well. In this way, we obtain
\be 
I(z)\ \sim\ \int d^4y\,S_{\mu}(z)\,\beta^{\mu}\,\mathscr{D}(y,z)\,
\big(-\partial_{\nu}J^{\nu}(y)\big)\ . \label{eq:cssc}
\ee
We see that the jets are connected to the soft part also by ``scalar
polarized'' gluons, which can be factored from the jets by using the
Ward identity given in Eq.~(\ref{ward1}). The
formal proof of factorization into hard, soft, and jet functions in
coordinate space now follows the momentum-space procedure, and
requires only the {\it hard-collinear} and {\it
soft-collinear} approximations described here.

\subsection{Eikonal approximation}\label{eik}

Having described the {\it hard-collinear} and {\it
soft-collinear} approximations above to factorize
the contributions from different subdiagrams at the leading
singularity, we can now think of another approximation to simplify the
computation of the leading term. One may make an
approximation to the integrals keeping only the leading contribution on
a pinch surface where the fermion lines are taken on the lightcone, and
neglecting the sub-leading contributions coming away
from that pinch surface by imposing the results of the Landau
conditions inside the integrands.

As an example, let us again take the fermionic vertex function. 
The solutions to Landau
equations with collinear fermions 
set the
transverse coordinates and the minus (plus) coordinates of the
positions of the fermion-gluon vertices on the plus-line
(minus-line) to zero as well as time-ordering them. These conditions
can be imposed inside the integrand by replacing the fermion
propagators along the plus line with 
\be S_F(x^2)\, =\,  \slashed{\partial}\,\Delta_F(x^2)\, \rightarrow\,
\theta(x^{+})\,\delta(x^{-})\,\delta^2(x_{\perp})\,\gamma\cdot\beta \ , \ee
with $\beta=\delta^{\mu +}$ while for those along the minus line, in
the direction $\beta'=\delta^{\mu -}$, $x^+$
and $x^-$ are exchanged. This is actually the coordinate-space version
of the well-known eikonal approximation, which is based on assuming
the gluons are soft and neglecting their squared momenta in the
fermion propagators. The eikonal
approximation originates from geometrical optics, where it corresponds
to the small wavelength limit in which the trajectories of light are given
by light rays as in classical theory. 
One might have
derived the form of the fermion propagators also by taking the Fourier
transform of the eikonal propagator in momentum space for a massless
fermion moving in the direction~$\beta^{\mu}=\delta^{\mu+}$,
\be \int \frac{d^4k}{(2\pi)^4}\,\frac{i}{\beta\cdot
  k+i\ep}\,e^{-ik\cdot x}\ =\ \theta(x^+)\,\delta(x^-)\,\delta^2(x_{\perp}) \ . \ee 
Let us apply this eikonal approximation to the one-loop vertex diagram
as an example, 
\be\begin{split} \Gamma^{(1)}_{\mathrm{eik}}\ =\ \int d^4y_2\,&d^4y_1\,
  \theta(x^+_2-y^+_2)\,\theta(y^+_2)\,\theta(y^-_1)\,\theta(x^-_1-y^-_1)
  \\
 &
 \ \times\delta(y^-_2)\,\delta^2(y_{2,\perp})\,\delta(y^+_{1})\,\delta^2(y_{1,\perp})\frac{1}{(-(y_2-y_1)^2+i\ep)}
 \ , \end{split}\ee
where we have suppressed the numerical factors and dropped the
delta-functions of the external lines. The result, after introducing
the parameters $\lambda$ and $\sigma$ with
$\beta'^{\nu}=\delta^{\nu-}$,
\be \Gamma^{(1)}_{\mathrm{eik}}\ =\ \int^{x^+_2}_0 d\lambda\int^{x^-_1}_0
d\sigma \,\, \frac{1}{(-2\beta\cdot\beta'\lambda\,\sigma+i\ep)} \ , \ee
is exactly equal to a first-order diagram of a Wilson line with a cusp
at the origin, which begins at the point $x^{\mu}_1$ {\it pointing} in the
direction of $\beta'$, then changes its direction to $\beta$ at the
origin, and later ends at~$x^{\mu}_2$\@. The parameters $\lambda$,
$\sigma$ are simply relabelings for $y^+_2$ and $y^-_1$ that give the
locations where the gluon is attached to the Wilson line, and of
course, are integrated over. This equality between the diagrams of a cusped path
ordered exponential and of the vertex function after the eikonal
approximation also holds at higher
orders, because the theta functions simply order the attachments to
the eikonals while the integrations over any other vertices are the
same in both cases. Therefore, we may approximate the vertex function by a
Wilson line calculation at any given order in perturbation
theory~\cite{Korchemsky:1987wg,Erdogan:2011yc}\@. The power counting
for the path ordered exponentials is not exactly the same with that
for the vertex function with partonic lines, but is very similar and
gives the same bound for their overall degree of divergence, which we
will present in Appendix~\ref{app:pcpoe} to avoid repetition.

\section{Discussion} 

The coordinate-space singularities of Feynman integrals in a massless
gauge theory have a direct interpretation in terms of physical
processes, in which classical massless
particles propagate freely between points in space-time, where they
scatter by local interactions. The singularities occur only if these
particles move on the lightcone. The condition for pinches in the
coordinate integrals is interpreted as momentum conservation for these
scattered particles with the identification of their momenta from
their coordinates. This interpretation is the same as the
interpretation given by Coleman and Norton~\cite{Coleman:1965xm} to
Landau equations in momentum space~\cite{Landau:1959fi}\@.

The pinches in the coordinate integrals for the vertex function occur
when a group of lines get mutually collinear forming jets as in momentum
space. There are also pinches from ``zero'' lines when some set of internal
vertices move to the origin $x\rightarrow 0$ reflecting a
short-distance singularity, where these zero lines or vertices with
vanishing components define a hard function in coordinate
space. There are also end-point singularities in the integrals over
Feynman parameters $\alpha\rightarrow 0$, which define the soft
function in coordinate space. An important difference from the
momentum space is that the soft function is finite in coordinate space
when the external points of the vertex function ($x_1$ and $x_2$
above) are kept at finite distances. The collinear
divergences are of ultraviolet nature in coordinate space, and require
$D<4$ in dimensional regularization; while no infrared regulation is
needed since the coordinates of the external particles provide the
natural infrared cut-off.

By the power counting arguments developed above, vertex
functions in coordinate space are found to be at worst logarithmically
divergent at higher orders, relative to the lowest order
results. Similarly, after the eikonal approximation, the path
ordered exponentials have the same bound for their
overall degree of divergence. The requirement for a divergence in
both cases is that the hard and soft subdiagrams must not be
directly connected, and they can only interact through the jets. Two jets
on the lightcone in different directions can only have a hard
interaction at the origin and interact softly at later times. This
illustrates in coordinate space the factorization of short and long-distance
dynamics in field theories. We have also explained the {\it hard-collinear} and
{\it soft-collinear} approximations that are needed for the formal
implementation of factorization in coordinate space.

The results of this study hold not only for a specific set of massless
gauge theories but are much more general. The results for vertex
functions can be generalized for fixed-angle scattering at large
angles, because there is no
interference between incoming and outgoing jets at large
angles~\cite{Aybat:2006mz,Sterman:2002qn,Catani:1998bh}\@. For
scattering at small angles, a different power counting for
the jets is needed. Furthermore, our discussion can be extended to S-Matrix
elements, defining the reduction from Green functions directly in
coordinate space, and eventually to cut diagrams for infrared-safe
cross-sections, topics which we will leave for future work.

\begin{acknowledgments}
The author is grateful to George Sterman for many helpful discussions
and suggestions. This
work was supported by the National Science Foundation Grants
No.~PHY-0969739 and No.~PHY-1316617. 
\end{acknowledgments}

\appendix
\section{Massive Lines}\label{app:mass}

For completeness, we consider the extension of pinch analysis to
massive lines in coordinate space. Massive lines must be explored
separately because the massive propagator in coordinate space has a
more complicated form, it can be written in $4-2\vep$~dimensions by
\be \Delta_F(x\,;\,m)\ =\ \left(\frac{-i\ }{8\pi^2}\right)\int^{\infty}_0
d\xi\,\left(\frac{2\pi
    i}{\xi}\right)^{\vep}\,\exp\left[i\left(-\frac{x^2}{2}\xi-\frac{m^2}{2\xi}+i\ep\right)\right]
\ .\label{master2} \ee
Since the massive propagator does not have a simple denominator, we
can not do a Feynman parametrization. However, using
Eq.~(\ref{master1}) we can combine the
propagators of each line of an arbitrary Feynman diagram with massive
lines, 
\be \tilde{I}(\{x^{\mu}_{i}\})= \prod_{\mathrm{lines}\
  j}\int^{\infty}_0d\xi_j\prod_{\mathrm{vertices}\
k}\int d^Dy_k \,\exp\Big[-i\tilde{D}(\xi_j,x_i,y_k)\Big]\times \tilde{F}(\xi_j,x_i,y_k) \ , \label{fgmaster2}
\ee
where $\xi_j$ is now the parameter of the
$j$th line with dimensions of mass square. The phase $\tilde{D}$ of
the exponent is given directly from (\ref{master2}) by
\be \tilde{D}(\xi_j,x_i,y_k)\ =\ \sum_j
\xi_j\frac{z_j^2}{2}+\frac{m_j^2}{2\xi_j} \ , \ee
with $z^{\mu}_j$ a linear function of the external coordinates
$\{x^{\mu}_i\}$ and the positions of (internal) vertices $\{y^{\mu}_k\}$
as before. The functions $\tilde{F}(\xi_j,x_i,y_k)$ include constants
and the ``numerators'', which might come from derivatives of three-point
vertices acting on the exponentials.

One can obtain the Landau conditions from the integral
representation (\ref{fgmaster2}) for diagrams with massive lines by
the method of stationary phase. The conditions of stationary
phase with respect to the positions of internal vertices,
\be   \frac{\partial}{\partial y^{\mu}_k}\,\tilde{D}(\xi_j,x_i,y_k)\ 
= \sum_{\mathrm{lines}\ j\ \mathrm{at\ vertex}\
  k}\eta_{jk}\,\xi_j\,z^{\mu}_j \ =\ 0 \ , \ee
where $\eta_{jk}=+1\ (-1)$ if the line~$j$ ends (begins) at
vertex~$k$, and is zero otherwise, 
give exactly the same result as Eq.~(\ref{landau2}) for the massless
case because the masses of the lines do not depend on the positions of
the internal vertices. The phase is stationary with
respect to the $\xi$-parameters when
\be  \frac{\partial}{\partial \xi_r}\,\tilde{D}(\xi_j,x_i,y_k)\ =\ 
 z^2_r\ -\ \frac{m^2_r}{\xi^2_r}\ =\ 0\ . \label{stat2} \ee
For massive lines, the stationary points are given by
\be \xi_r\ =\ \frac{m_r}{\sqrt{z^2_r}}\ ,\quad\quad z^2_r>0 \ . \label{stat2b}\ee
If the mass of
the line $r$ is zero, Eq.~(\ref{stat2}) is only satisfied if its
coordinates have a lightlike separation, irrespective of the value of
$\xi_r$\@.

Repeating the same reasoning as in the massless case, we
identify the product $\xi_rz^{\mu}_r$ with a momentum vector
$p^{\mu}_r$ for line $r$, while this time $\xi_r$ is determined by
(\ref{stat2b}). The time-component of this momentum vector,
\be p^0_r\ =\ 
m_r\,\frac{z_r^0}{\sqrt{(z_r^0)^2-|\vec{z}_r|^2}}\ =\
\frac{m_r}{\sqrt{1-\beta^2_r}}\ , \ee
equals the energy of a classical particle with mass $m_r$ propagating
with the speed of $\beta_r=|\vec{z}_r|/z^0_r$\@. Therefore, we
can interprete the stationary phases in the integral
representation (\ref{fgmaster2}), in the same way as pinch
singularities explained in Sec.~\ref{analysis}, as a physical process in
space-time where classical particles
propagate between vertices with their momenta conserved at each
vertex.

\section{Power Counting for Path Ordered
  Exponentials}\label{app:pcpoe}

We shall lastly do the power counting for the vacuum expectation value
of path ordered exponentials with constant lightlike velocities meeting at a
cusp. Consider one Wilson line, which starts from the point
$x^{\mu}_1=x_1\,\delta^{\mu-}$ in
\mbox{$\beta_1^{\mu}=\delta^{\mu-}$} direction, and meets the other line at
the origin, which moves in
$\beta^{\mu}_2=\delta^{\mu+}$ direction and ends at the point
$x^{\mu}_2=x_2\,\delta^{\mu+}$\@. Formally, we consider the diagrams
for the vacuum expectation value of the following operator,
\be
\Gamma_{\beta_1,\beta_2}(x_1,x_2)
=
\bigg \langle 0\left| T\bigg(  \Phi_{\beta _2}(x_2,0)\,
  \Phi_{\beta _1}(0,x_1) \bigg)\right|0  \bigg \rangle\  ,\ee
with constant-velocity ordered exponentials,
\be
\Phi_{\beta_i}(x+\lambda\beta_i,x) = {\cal P}\exp \left (-ig\int_0^\lambda d\lambda'\beta _i\cdot A(x+\lambda'\beta_i)
\right )\  . \ee
As for the
vertex function, there may
be divergences when some vector and/or fermion lines get collinear to
the eikonal lines forming two jets, which can interact softly at large
distances and have a hard interaction at the cusp.

In analogy to Eq.~(\ref{gammaoverall}) for the vertex function, the overall
degree of divergence of such path ordered exponentials in coordinate
space can be written with a bound from below,
\bea
\gamma^{\mathrm{eik}} &\ \geq \ &
w^H+4(V^H_3+V^H_4)-2N^H_g-3N^H_f-V_{3g}^H \label{gammaeik}  \nn \\
     & & +\sum_{i=+,-} \Big[
     2(V^{J_{(i)}}_3+V^{J_{(i)}}_4)-N^{J_{(i)}}_g-2N^{J_{(i)}}_f-V^{J_{(i)}}_{3g} \\
 & & \quad\quad\quad\ \, +\frac{1}{2}(V_{3}^{J_{(i)}}-S^J_{(i)}-J^H_{(i)g}+w^{J_{(i)}})\Big]\ , \nn 
\eea
with $w^H$ the total number of hard lines attached to the Wilson
lines, and $w^{J_{(\pm)}}$ the number of attachments of the jet in the
$\pm$~direction to the Wilson line in the same
direction. The lines that connect a jet to the Wilson line in the
opposite direction are soft
lines, and are counted together with the connections of the jet to the soft
subdiagram by $S^J_{(i)}$\@.

In Eq.~(\ref{gammaeik}), we have 
added a term $+w^H$ for the integrations over the
locations of the attachments of the hard subdiagram to the eikonals to the
contributions from the hard part, because these connections have
to move to the cusp in
order that all components of these hard lines vanish. Furthermore, the
derivatives at each three-gluon vertex will bring vectors that
form invariants in the numerator, which will be of the form
either $(\beta_i\cdot z)$ or $(z\cdot z')$\@. The number of vectors~$\beta_i$ 
that can show up in the numerator is equal to the sum of the number of
attachments to each Wilson line. The net effect from all three-gluon
vertices in the hard subdiagram is given by the term $-V_{3g}^H$ as
for the vertex function. In a jet, a three-gluon vertex~$z$ 
that is connected to the Wilson line in the same direction as the jet
produces an invariant $(\beta_i\cdot
z)$ linear in $\lambda$ in the numerator, while one connected to the
opposite Wilson line produces an invariant zeroth order
in~$\lambda$\@. Therefore, we add $w^J=w^{J_{(+)}}+w^{J_{(-)}}$ to the
number of jet three-point vertices for the term for the minimum numerator
suppressions in $\gamma^{\mathrm{eik}}$, and 
subtract the connections of the jets to the opposite eikonals with
those to the soft subdiagram.

The relations of the
number of lines to the number of the vertices for the jets and the
hard subdiagram are in
this case slightly different than Eqs.~(\ref{hardnumber}) and
(\ref{jetnumber}) for the vertex function,
\bea 
2N^H+J^H+S^H& = & w^H + 3V^H_{3} + 4V^H_4 \ , \\
2N^J+S^J& = & w^J + J^H + 3V^J_{3} + 4V^J_4 \ , 
\eea
for the hard part and the jets, respectively. Similarly, the relation
between the number of fermion lines and the fermion-gluon vertices in
the hard part is given by
\be V^H_{3f} = N^H_f +\frac{1}{2}(S^H_f+J^H_f) \ , \ee
while for the jets they are related by
\be V^J_{3f} = N^J_f +\frac{1}{2}(S^J_f-J^H_f) \ . \ee
Note also that $J^H_f\geq 0$ in this case. Plugging these graphical
identities for the subdiagrams into Eq.~(\ref{gammaeik}), we find
\be \gamma^{\mathrm{eik}}\ \geq \ S^H_g + \frac{3}{2}S^H_f +\frac{1}{2}(S^J_f+J^H_f) \ . \ee
Any direct connection between the hard and soft subdiagrams and
fermion lines connecting any two subdiagrams suppress
the integral as for the vertex function. The collinear singularities
of path ordered exponentials with constant lightlike velocities are
also at worst logarithmic in coordinate
space~\cite{Korchemskaya:1992je}\@.

%%%%%%%%%%%%%%%%%%%%%%%%%%%%%%%%
%% References 
%%%%%%%%%%%%%%%%%%%%%%%%%%%%%%%%

%%%%%%%%%%%%%%%%%%%%%%%%%%%%%%

\end{document}